\documentclass[11pt]{article}

\usepackage{amsmath}
\usepackage{amsfonts}
\usepackage{stix}
\usepackage{hyperref}

\usepackage[margin=0.8in]{geometry}
\usepackage{graphicx}
\usepackage{caption}
\usepackage[labelformat=parens,labelfont={}]{subcaption}

\usepackage{floatrow}
\newfloatcommand{capbtabbox}{table}[][\FBwidth]

\usepackage{blindtext}

\begin{document}

\title{Machine learning materials physics: Deep neural networks trained on elastic free energy data from martensitic microstructures predict homogenized stress fields with high accuracy}
\author{K. Sagiyama\thanks{Mechanical Engineering, University of Michigan} and K. Garikipati\thanks{Mechanical Engineering, Mathematics, and Michigan Institute for Computational Discovery \& Engineering, University of Michigan; corresponding author: {\tt krishna@umich.edu}}}
\date{December 31, 2018}

\maketitle
\abstract{
We present an approach to numerical homogenization of the elastic response of microstructures. Our work uses deep neural network representations trained on data obtained from direct numerical simulation (DNS) of martensitic phase transformations.  The microscopic model leading to the microstructures is based on non-convex free energy density functions that give rise to martensitic variants, and must be extended to gradient theories of elasticity at finite strain. These strain gradients introduce interfacial energies as well as coercify the model, enabling the admission of a large number of solutions, each having finely laminated microstructures. The numerical stiffness of these DNS solutions and the fine scales of response make the  data expensive to obtain, while also motivating the search for homogenized representations of their response for the purpose of engineering design. The high-dimensionality of the problem is reduced by training deep neural networks (DNNs) on the effective response by using the scalar free energy density data. The novelty in our approach is that the trained DNNs also return high-fidelity representations of derivative data, specifically the stresses. This allows the recapitulation of the classic hyperelastic response of continuum elasticity via the DNN representation.  Also included are detailed optimization studies over hyperparameters, and convergence with size of datasets.
}

\section{Introduction}
Martensite is a very hard crystalline structure that is formed by diffusion-less transformations.  
In this work we focus on martensites resulting from cubic-tetragonal martensitic phase transformations, which include many important industrial materials such as barium titanate $\textrm{BaTiO}_3$ in capacitors, lithium manganese dioxide $\textrm{LiMnO}_2$ in battery electrodes, shape-memory alloys, and steel.
Better understanding of its formation and behaviour is crucial to better material designs; 
the hardness of steel, for instance, depends on its martensite content, too much martensite making steel brittle, while too little making it soft.  

Mathematical formulations and direct numerical simulations (DNSs) of the martensites in this category have been studied \cite{Rudraraju2014,Sagiyamaetal2016,Rudrarajuetal2016,SagiyamaGarikipati2017b}, addressing the formation of martensite in the context of continuum mechanics, where gradient-coercified hyperelasticity at finite strain was solved under non-convex free energy density functions using isogeometric analysis \cite{Cottrell2009}.  
For practical material designs, however, such approaches assume computational expense that is prohibitive if a large number of designs are to be tested, because accurate DNS would require spatial meshes fine enough to resolve the interfaces between tetragonal phases.  
In addition \emph{microscopic} details such as phase distributions over a volume are often of little practical interest, and fast computation of \emph{macroscopic} material behaviour, such as effective/homogenized stress-strain relations, is the key to accelerate design and discovery of new materials.  
To this end, it is of practical importance to develop homogenized models of those multi-phase martensites, and our goal in this work is to develop such models using \emph{data} sets directly generated by DNS.

Quite a few attempts have been made at data-driven homogenization of different types of microstructures, as well as of larger scale structres.  
The problem requires data sets of descriptors (or inputs), such as strain, and quantities of interest (or outputs), such as stress, which can be obtained from experiments as well as from DNSs.  
These data sets are then used to form a general representation of the output quantities in terms of the input features.  
Nonlinearly elastic heterogeneous materials under finite/infinitesimal strain have been studied in a series of works  \cite{Yvonnet2009,Clement2012,Yvonnet2013,Le2015}.
Representations of effective potentials in terms of nine input features, such as the macroscopic strains and the volume fractions, have been sought for nonlinearly elastic heterogeneous materials with the infinitesimal strain assumption using neural networks (NNs)  \cite{Le2015}.  
Solutions were computed at randomly selected points in the input feature space using finite element methods (FEM), where a \emph{fixed-point algorithm} previously proposed  \cite{Manzhos2010} was used for lower computational and memory requirements.  
There, the ability of NNs to represent high-dimensional input was successfully demonstrated.  
A comprehensive study of homogenization of hyperelastic and inelastic composites has been presented \cite{Bessa2017}, where,
for two-dimensional heyperelastic composites, quantities such as strain components, particle volume fraction, and particles' semi-axis aspect ratio were used as input features and the effective free energy density and macroscopic stress components were quantities of interest.  
Inputs were sampled based on Sobol' sequences in the feature space, and the output quantities were computed using FEM.  
For three-dimensional inelastic composites, two quantities related to yielding and hardening of particles were used as inputs and material toughness was used as output. 
Those inputs were again sampled using a Sobol' sequence \cite{Sobol1967} in the two-dimensional input variable space.  
Kriging and neural network architectures were used as mathematical models that relate inputs and outputs, and promising homogenized material laws were successfully found.  
Inelastic steel composite materials also have been considered \cite{Gupta2015}.  
Various two-dimensional microstructures with randomly chosen parameters were created and their mechanical responses such as the effective (composite) yield strength were numerically computed using FEM.
\emph{N-point statistics} with $N=2$ was used along with principal component analysis to systematically extract two significant combinations of input features, whose corresponding eigenvalues were then used as input features in the subsequent polynomial regression analysis to predict macroscale parameters of interest.
The same approach was also used for effective diffusivity of the porous transport layers in a Polymer Electrolyte Fuel Cell (PEFC)  \cite{Cecen2014}, and for the effective elastic stiffness components of a porous elastic solid \cite{Kalidindi2015} .
On the other hand,  a \emph{data-driven computing} paradigm that does not rely on mathematical representation of constitutive relations also has been proposed \cite{Kirchdoerfer2016}.    
There, with kinematic compatibility and equilibrium being constrained, constitutive relations were imposed by direct use of experimental material data points.  
This novel paradigm was applied to examples of nonlinear elastic trusses  and linear elasticity.

In this work we consider computational homogenization of three-dimensional twin microstructures resulting from martensitic phase transformations, whose formation is modeled by finite strain gradients coercifying hyperelasticity with non-convex free energy density functions.  
Our overarching goal is to find mathematical representations of the macroscopic constitutive laws with input features being macroscopic strains and microstructural details and output quantities being the effective free energy density and the macroscopic stresses.  In this initial communication, however, we work with a single microstructure, and postpone the incorporation of microstructural features to the near future.
We sample input features using Sobol' sequences as previously introduced to the field \cite{Bessa2017}, and numerically compute solutions at each point using isogeometric analysis \cite{Cottrell2009}.   
We use NNs, as others have done \cite{Le2015,Bessa2017}, as the mathematical model architecture to represent the macroscopic constitutive laws of interest.  
To our knowledge, this is the first work on computational homogenization that deals with twin microstructures.

In Sec.~\ref{sec:homogmeth} we layout the homogenization procedure and assess the validity of using DNNs for homogenization of the elastic response by  using the neo-Hookean model of hyperelasticity to generate DNS data. With that as a guideline, we turn to the problem of DNN representation for the homogenization of martensitic microstructures formed using a gradient-coercified model of elasticity at finite strain and non-convex free energy in Sec~\ref{sec:ge}. Discussion and conclusions follow in Sec~\ref{sec:conclusion}.

\section{Homogenization methodology for a neo-Hookean material}
\label{sec:homogmeth}
Throughout this study, we consider materials on a reference unit cube $\Omega = (0,1)^3$ with periodic boundary conditions \cite{Yvonnet2009,Yvonnet2013,Le2015,Bessa2017}.  In this section we assume that the material's response is given by the neo-Hookean hyperelastic free energy density function. Suppose that  $\Omega$ is subject to boundary conditions under which it has the average Green-Lagrange strain $\overline{\boldsymbol{E}}$. From it we define the right Cauchy-Green tensor $\overline{\boldsymbol{C}}=2\overline{\boldsymbol{E}}+\boldsymbol{I}$, where $\boldsymbol{I}$ is the second-order isotropic tensor, and consider a polar decomposition of $\overline{\boldsymbol{C}}=\boldsymbol{\Phi}^\textrm{T}\boldsymbol{\Lambda}^2\boldsymbol{\Phi}$.  
We define the average deformation gradient  $\overline{\boldsymbol{F}}=\boldsymbol{\Phi}^\textrm{T}\boldsymbol{\Lambda}\boldsymbol{\Phi}$. The current (deformed) position of material points is
\begin{align*}
\boldsymbol{x}=\overline{\boldsymbol{F}}\boldsymbol{X}+\boldsymbol{u}.
\end{align*}
where $\boldsymbol{u}$ accommodates periodic boundary conditions. Under this mode of deformation, the true deformation gradient is $\boldsymbol{F} = \overline{\boldsymbol{F}} + \partial\boldsymbol{u}/\partial\boldsymbol{X}$, where $\boldsymbol{I}$ is the identity tensor. With the free energy density function $\Psi(\boldsymbol{F})$ and we compute the first Piola-Kirchhoff stress $\boldsymbol{P} = \partial\Psi/\partial\boldsymbol{F}$.

\subsection{Microscopic model}\label{E:nh_micro}

The ``microscopic'' model is simply the neo-Hookean free energy density function,  

\begin{align}
    \Psi = \frac{\mu}{2}(I_1 -3 -2\ln J) + \frac{\lambda}{2} (\ln J)^2,
\end{align}
where $I_1$ is the first invariant of the right Cauchy-Green tensor $\boldsymbol{C} = \boldsymbol{F}^\mathrm{T}\boldsymbol{F}$, or $I_1=C_{11}+C_{22}+C_{33}$, and $J$ is the determinant of $\boldsymbol{C}$. The secon Piola-Kirchhoff stress is then,
\begin{align}
    \boldsymbol{P} = \mu \boldsymbol{F}(\boldsymbol{I}-\boldsymbol{C}^{-1}) + \lambda (\ln J) \boldsymbol{F}\boldsymbol{C}^{-1}.
\end{align}

All computations were carried out using isogeometric analysis (IGA) \cite{Cottrell2009} within the mechanoChem library available at \url{https://github.com/mechanoChem/mechanoChem}.
\subsection{Macroscopic modeling using DNNs}\label{SS:nh_macro}

The periodic boundary conditions give:
\begin{align}
\boldsymbol{F}_\mathrm{avg} &= \int\limits_\Omega \boldsymbol{F} \mathrm{d}V\label{eq:Favg}\\
\overline{\Psi}(\overline{\boldsymbol{E}}) &:=  \int\limits_\Omega \Psi(\boldsymbol{F}) \mathrm{d}V\label{eq:barpsi}\\
 \overline{\boldsymbol{S}}&:=\overline{\boldsymbol{F}}^{-1}\overline{\boldsymbol{P}} \mathrm{d}V\label{eq:barS}
\end{align}
where $\overline{\boldsymbol{S}}, \overline{\boldsymbol{P}}$ are, resepctively, the macroscopic second and first Piola-Kirchhoff stress tensors, and $\overline{\boldsymbol{P}}$ is computed from the surface averages of the traction components on a given surface of $(0,1)^3$ with normal $\boldsymbol{N}$ in the positive/negative $J^\mathrm{th}$ direction.
\begin{equation}
\overline{P}_{iJ} = \int\limits_\Gamma P_{iK}N_K \mathrm{d}A_J.
\label{eq:barP}
\end{equation}

\subsubsection{Data sampling}\label{SSS:nh_macro_data}

\begin{figure}
\begin{center}
        \includegraphics[scale=0.35]{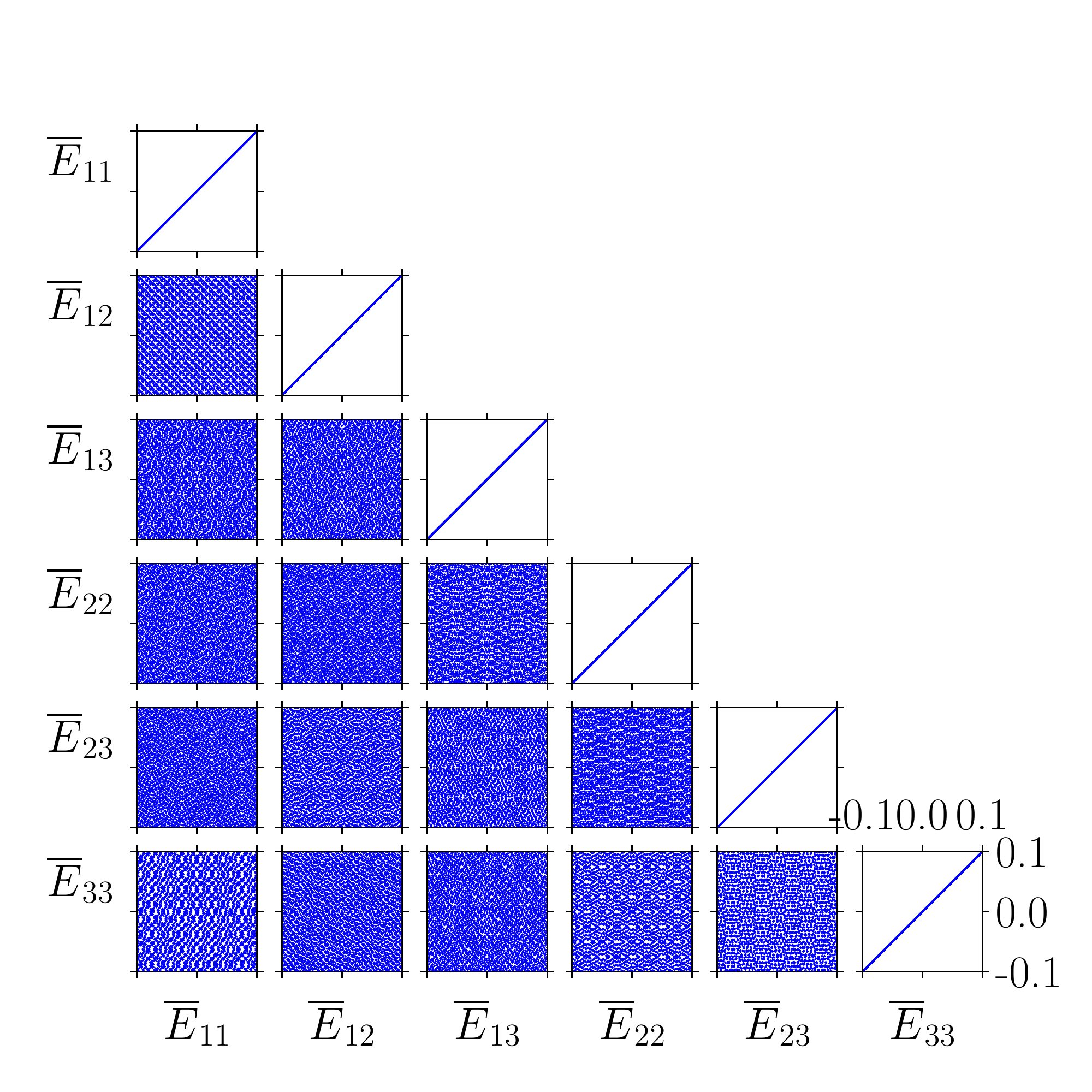}
    ~
\end{center}
\caption{Distribution of sample data points based on a Sobol' sequence for $\overline{\boldsymbol{E}} \in \mathbb{R}^6$: 4096 points were generated in $[-0.1,0.1]^6$ to obtain data for $\overline{\Psi}$ and $\overline{\boldsymbol{S}}$ from the neo-Hookean hyperelastic free energy density function.  Projections of these points onto the two-dimensional hyperplanes $\overline{E}_{IJ}-\overline{E}_{I'J'}$ are shown.  }
 \label{Fi:nh_dens}
\end{figure}

We generated a Sobol' sequence with 4096 inputs for $\overline{\boldsymbol{E}}$ sampled as  points $\{\overline{E}_{11}, \overline{E}_{22}, \overline{E}_{33}, \overline{E}_{12}, \overline{E}_{23}, \overline{E}_{13}\} \in [-0.1,0.1]^6$ using the GNU Scientific library \cite{gsl}; those points projected onto the two-dimensional hyperplanes $E_{IJ}-E_{I'J'}$ are shown in Fig. \ref{Fi:nh_dens}.  
No pair of two points falls to a single point upon these projections and, more generally, upon a projection to any $E_{IJ}$-axis.

\subsubsection{DNN representations}

We gather data sets $\{\overline{\Psi},\overline{\boldsymbol{S}}\}$ for each $\overline{\boldsymbol{E}}$. Next, we train a DNN against these data for $\overline{\Psi}$ with inputs $\overline{\boldsymbol{E}}$, and denote by $\overline{\boldsymbol{\Psi}}_\mathrm{NN}(\overline{\boldsymbol{E}})$ the response of the machine learned representation. See Sec~\ref{SSS:ge_macro_single_NN} for details on our DNN formulation, laid out there in the context of the gradient-coercified non-convex hyperelasticity model at finite strain that gives rise to martensitic microstructure. Using the DNN representation $\overline{\boldsymbol{\Psi}}_\mathrm{NN}(\overline{\boldsymbol{E}})$ we compute the DNN prediction for the second Piola-Kirchhoff stress 
\begin{equation}
\overline{\boldsymbol{S}}_\mathrm{NN}= \frac{\partial\overline{\Psi}_\mathrm{NN}}{\partial\overline{\boldsymbol{E}}}
\label{eq:DNN_hyperelas}
\end{equation}


\subsection{Validation of the DNN representation}\label{SSS:nh_macro_NN}

\begin{figure}
\begin{center}
        \begin{subfigure}[b]{3in}
            \centering
            \includegraphics[scale=0.35]{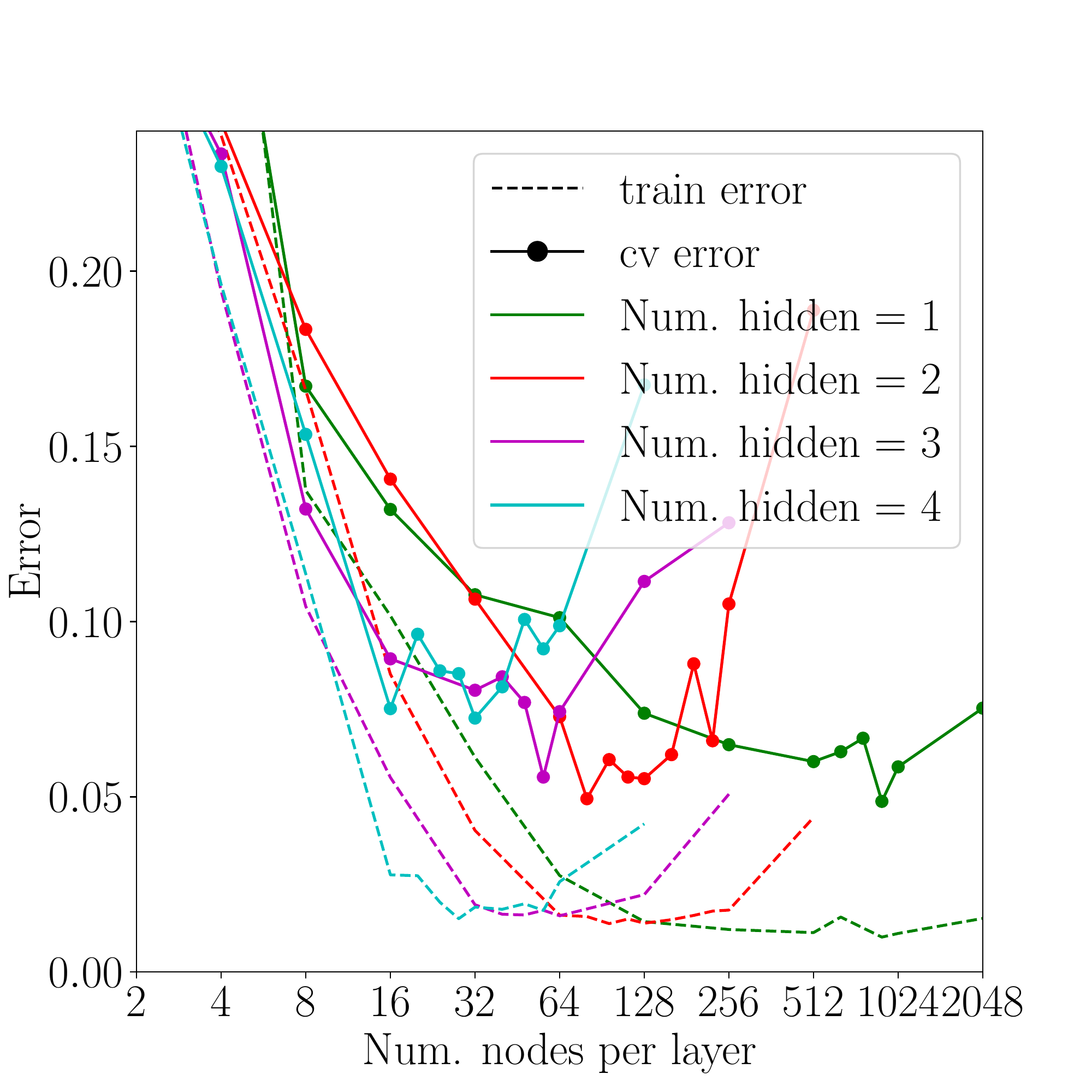}
            \caption{}
            \label{Fi:nh_cv_32}
        \end{subfigure}
        ~
        \begin{subfigure}[b]{3in}
            \centering
            \includegraphics[scale=0.35]{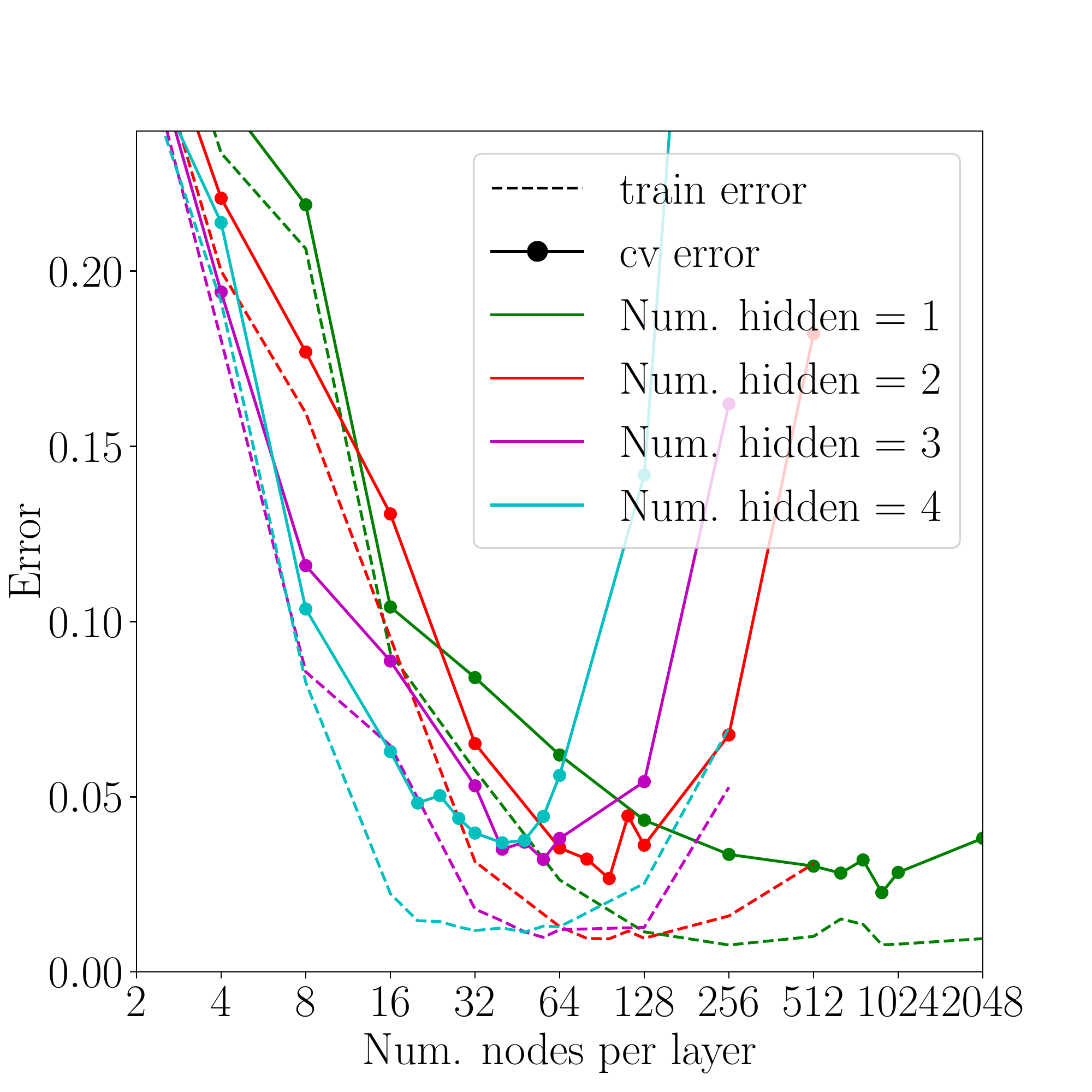}
            \caption{}
            \label{Fi:nh_cv_64}
        \end{subfigure}
        
        \begin{subfigure}[b]{3in}
            \centering
            \includegraphics[scale=0.35]{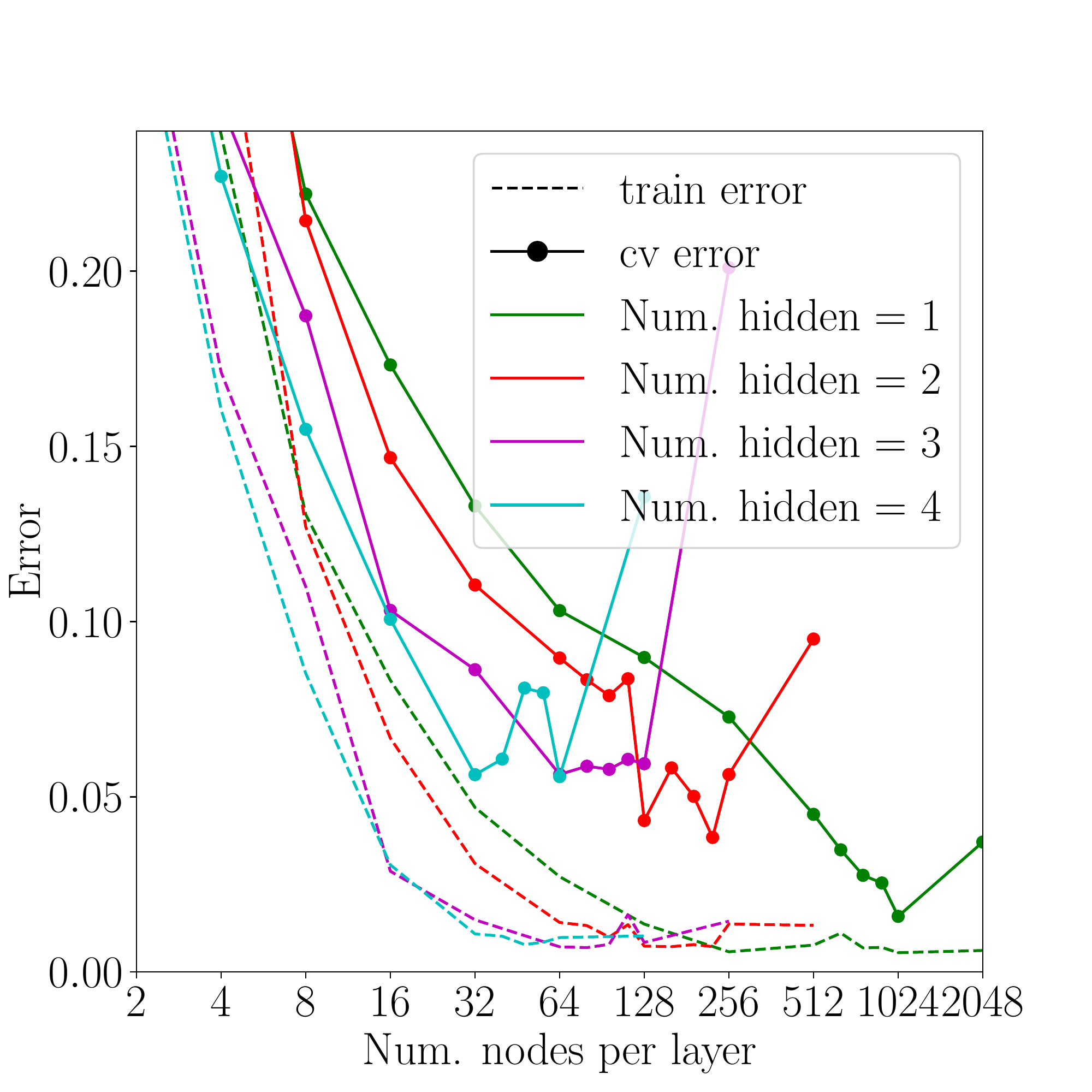}
            \caption{}
            \label{Fi:nh_cv_128}
        \end{subfigure}
        ~
        \begin{subfigure}[b]{3in}
            \centering
            \includegraphics[scale=0.35]{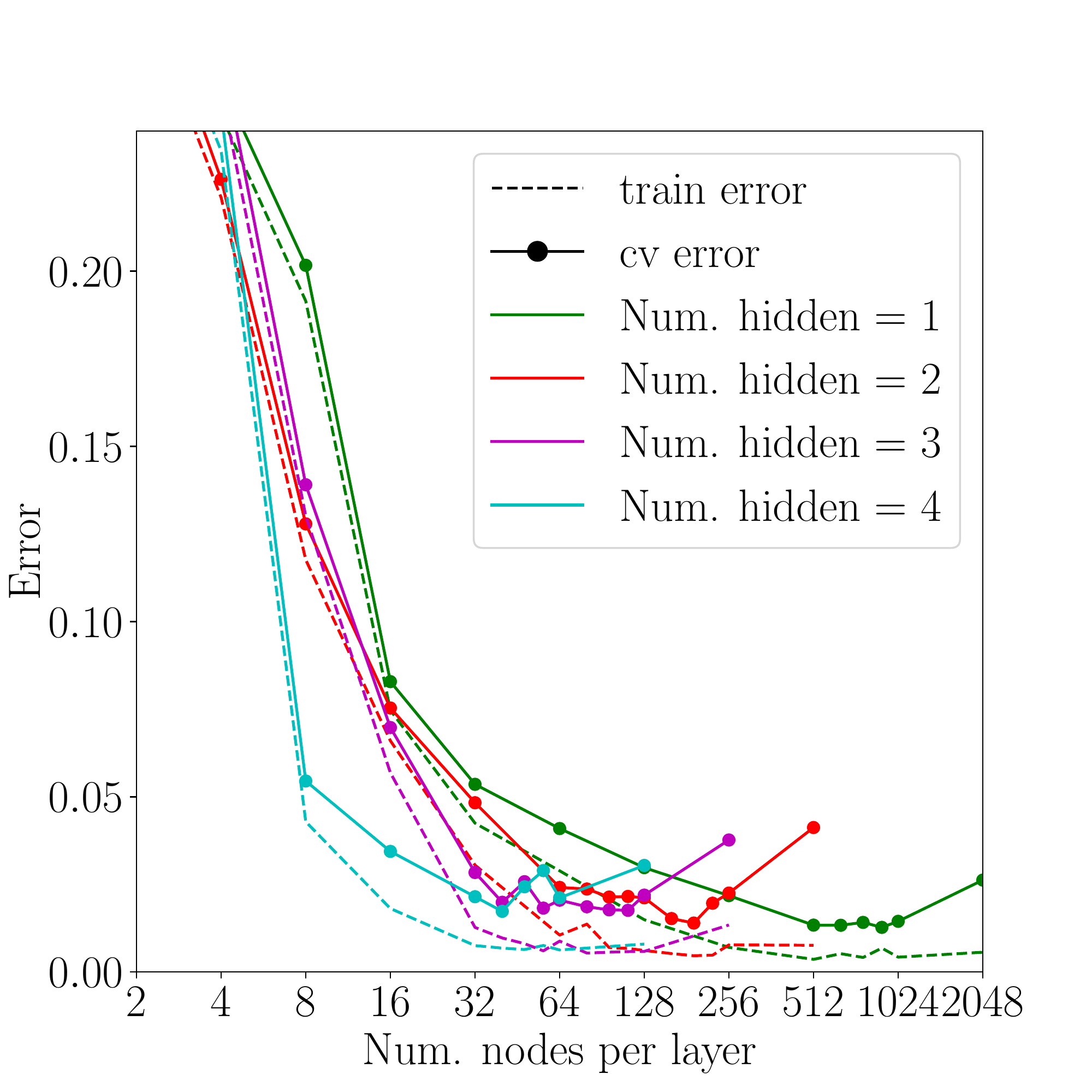}
            \caption{}
            \label{Fi:nh_cv_256}
        \end{subfigure}
\caption{Training and cross-validation errors in $\overline{\Psi}_\mathrm{NN}$ for varying dataset sizes: (\subref{Fi:nh_cv_32}) $N=256$, (\subref{Fi:nh_cv_64}) $N=512$, (\subref{Fi:nh_cv_128}) $N=1024$, and (\subref{Fi:nh_cv_256}) $N=2048$ for data from the neo-Hookean hyperelastic free energy density function.  Errors are plotted in log-scale for different number of hidden layers, $H=1,2,3,4$, and various number of nodes per hidden layer, $O$, between 2 and 2048.  }
\label{Fi:nh_cv}
\end{center}
\end{figure}

\begin{figure}
\begin{center}
        \begin{subfigure}[b]{3in}
            \centering
            \includegraphics[scale=0.35]{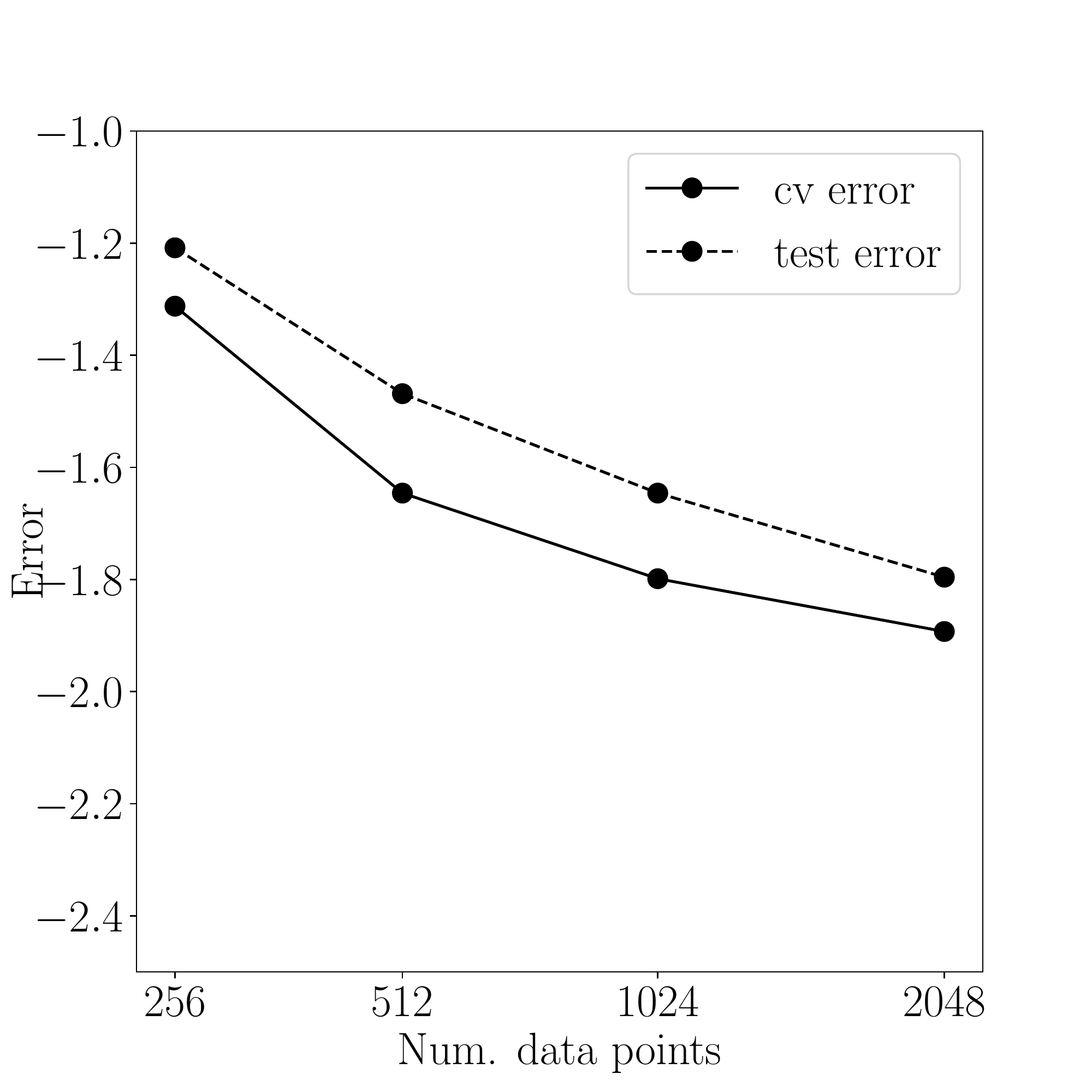}
            \caption{}
            \label{Fi:nh_Uconv}
        \end{subfigure}
        ~
        \begin{subfigure}[b]{3in}
            \centering
	    \includegraphics[scale=0.35]{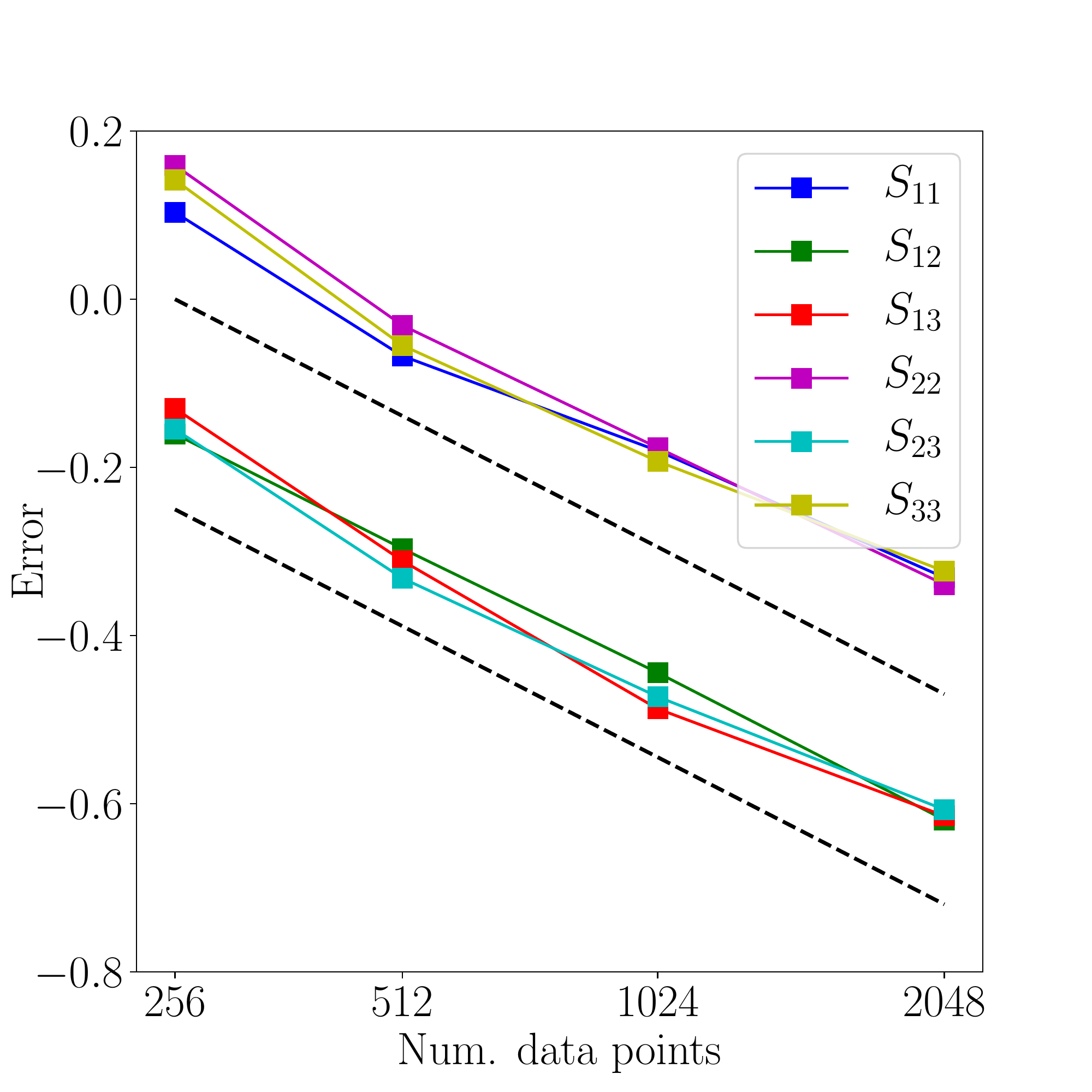}	
            \caption{}
            \label{Fi:nh_Sconv}
        \end{subfigure}

\caption{Convergence of neural network predictions to the \emph{correct} DNS data with respect to data resolution for (\subref{Fi:nh_Uconv}) $\overline{\Psi}_\textrm{NN}$ and (\subref{Fi:nh_Sconv}) $(\overline{S}_{IJ})_\textrm{NN}$ with the neo-Hookean strain energy density function.  Cross-validation and test errors were separately computed for $\overline{\Psi}_\textrm{NN}$ and all data were used for $\overline{\boldsymbol{S}}_\textrm{NN}$.}
\end{center}
\end{figure}

Figures \ref{Fi:nh_cv_32}-\ref{Fi:nh_cv_256} show the results of optimization of DNN hyper parameters by training and cross validation against data generated by  the neo-Hookean material model.  The cross validation and test results are summarized in Table \ref{Ta:nh_cv}. Figure \ref{Fi:nh_Uconv} shows the learning curves for the DNN representation of  $\overline{\Psi}_\mathrm{NN}$ as the training dataset size, $N$ increases. Each point on these curves is obtained from the corresponding hyper parameter optimization curve in Figure \ref{Fi:nh_cv} for a single hidden layer and the optimal number of nodes taken from Table \ref{Ta:nh_cv}. Figure \ref{Fi:nh_Sconv} shows the error between the ``macroscopic'' stress components $\overline{\boldsymbol{S}}_\mathrm{NN}$ obtained from the DNN using Equation \eqref{eq:DNN_hyperelas} and the stress data   $\overline{\boldsymbol{S}}$ from the neo-Hookean model, Equation \eqref{eq:barS}. The dashed lines show that this error scales as $\sim N^{-1/6}$, with the one-sixth power reflecting the volume of the strain space, with $\overline{\boldsymbol{E}} \in \mathbb{R}^6$. This is a notable result showing that the DNN trained to $\overline{\Psi}_\mathrm{NN}(\overline{\boldsymbol{E}})$ also delivers high-fidelity predictions via derivative quantities, $\overline{\boldsymbol{S}}_\mathrm{NN}$ in Equation \eqref{eq:DNN_hyperelas}. This is the result that we seek to generalize for DNN representations of the homogenized hyperelastic response of martensitic microstructures formed from non-convex free energy density functions, in Sec \ref{sec:ge}.

\begin{table}
  \begin{tabular}{|c|c|c|c|c|}
    \hline
    N & H & O & cv & test\\
    \hline
    256 & 1 & 896 & 0.0487 & 0.0619\\
    \hline
    512 & 1 & 896 & 0.0226 & 0.0340\\
    \hline
    1024 & 1 & 1024 & 0.0159 & 0.0226\\
    \hline
    2048 & 1 & 1024 & 0.0128 & 0.0164\\
    \hline
  \end{tabular}
  \caption{Optimal number of hidden layers $H$ and number of nodes per hidden layer $O$ for training dataset size $N=256,512,1024,2048$ from Fig. \ref{Fi:nh_cv}; corresponding cross-validation and test errors are also shown for the case of the neo-Hookean hyperelastic free energy density function}%
  \label{Ta:nh_cv}
\end{table}

\begin{figure}
\begin{center}
        \begin{subfigure}[b]{3in}
            \centering
            \includegraphics[scale=0.35]{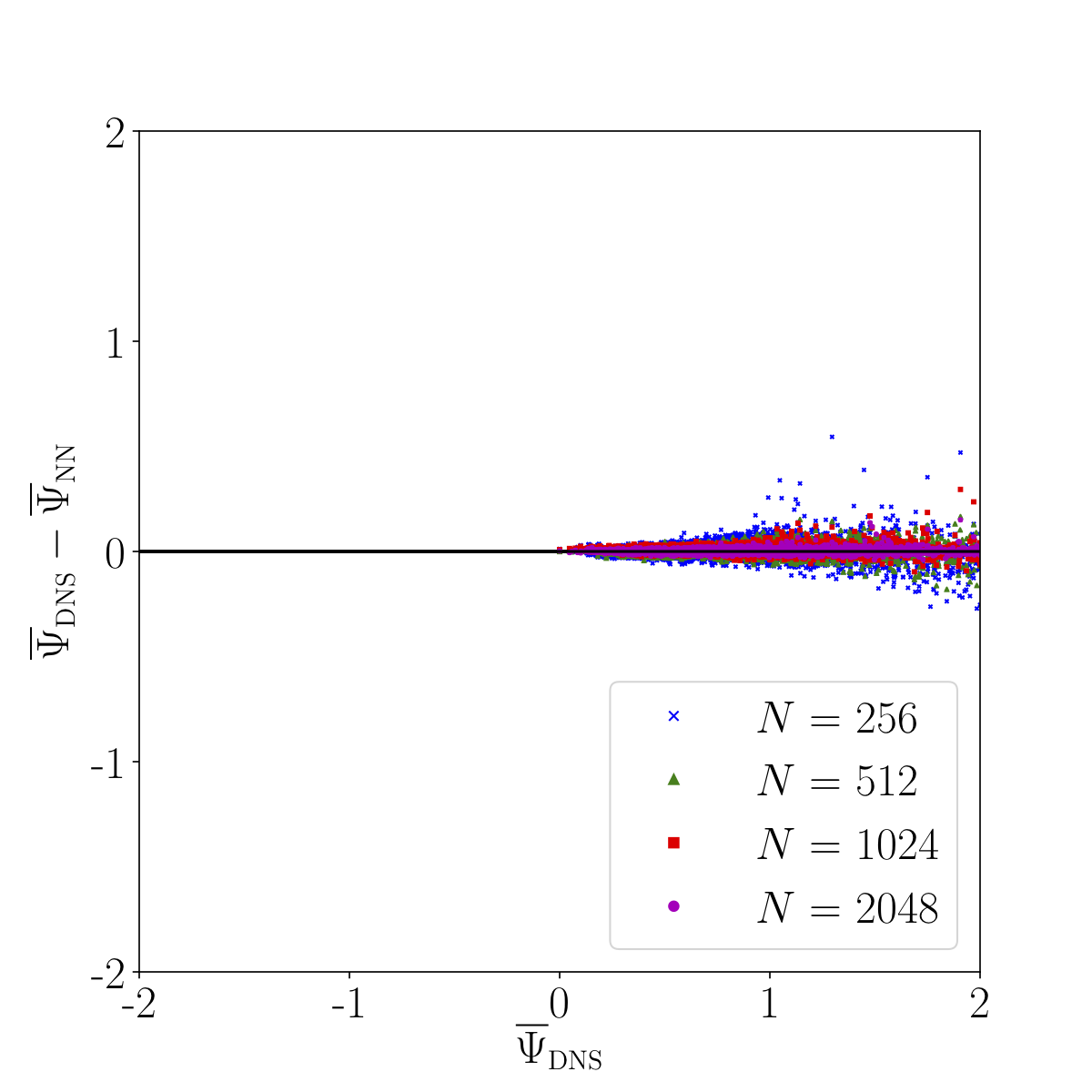}
            \caption{}
            \label{Fi:nh_Uerror}
        \end{subfigure}
        ~
        \begin{subfigure}[b]{3in}
            \centering
            \includegraphics[scale=0.35]{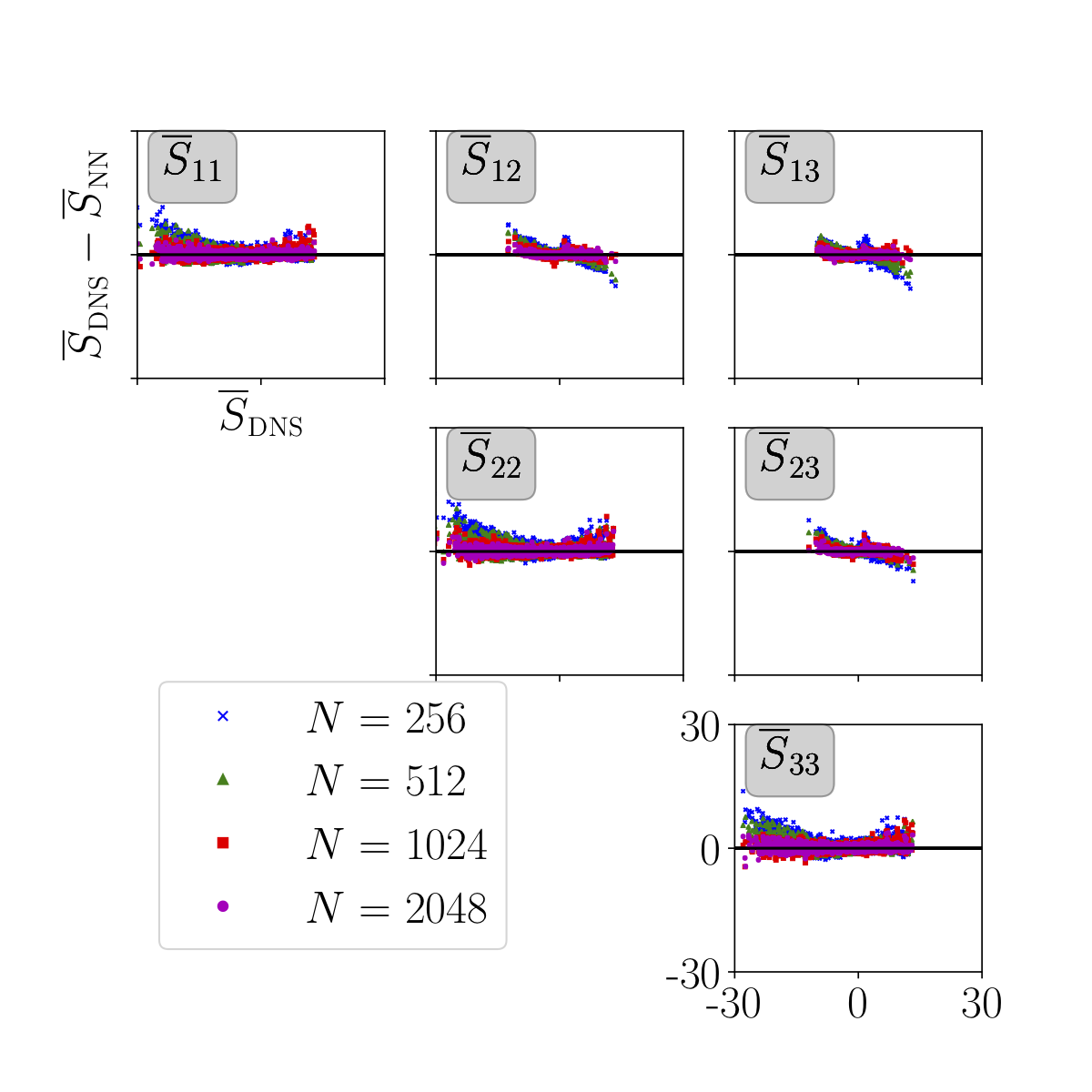}
            \caption{}
            \label{Fi:nh_Serror}
        \end{subfigure}
\caption{Absolute error of the neural network predictions when compared to the \emph{correct} values from DNS for (\subref{Fi:nh_Uerror}) $\overline{\Psi}$ and (\subref{Fi:nh_Serror}) $\overline{S}_{IJ}$ for the case of the neo-Hookean hyperelastic free energy density function.  Neural network predictions are computed and are compared to DNS data for all 4096 data points using the optimal network structures obtained in Table \ref{Ta:nh_cv} for each $N$ .  }
\end{center}
\end{figure}

The final result for the neo-Hookean material model and its DNN representation is in Figures \ref{Fi:nh_Uerror} and \ref{Fi:nh_Serror}, which show the absolute errors in $\overline{\Psi}_\mathrm{NN}$ and $\overline{\boldsymbol{S}}_\mathrm{NN}$. We note that the DNN representation loses some fidelity for larger magnitudes of free energy and the stresses, i.e., as the nonlinearity of response increases.

\section{DNN homogenization of martensitic microstructures obtained from gradient-coercified  non-convex hyperelasticity at finite strain} \label{sec:ge}

Proceeding to the problem of interest, that of developing DNN representations for numerical homogenization of the hyperelastic response of martensitic microstructures, we first present the underlying ``microscopic'' model.

\subsection{The microscopic model of gradient-coercified non-convex hyperelasticity at finite strain with general boundary conditions}
We solve for the displacement field $\boldsymbol{u}$ in $\Omega$.  
In this section we assume that $\boldsymbol{u}$ and its spatial derivatives are continuously defined in $\overline{\Omega}$.  
The boundary of $\Omega$ is assumed to be decomposed into a finite number of smooth surfaces $\Gamma_{\iota}$, smooth curves $\Upsilon_{\iota}$, and points $\Xi_{\iota}$, so that $\partial\Omega=\Gamma\cup\Upsilon\cup\Xi$ where $\Gamma=\cup_{\iota}\Gamma_{\iota}$, $\Upsilon=\cup_{\iota}\Upsilon_{\iota}$, and $\Xi=\cup_{\iota}\Xi_{\iota}$.  
Each surface $\Gamma_{\iota}$ and curve $\Upsilon_{\iota}$ is further divided into mutually exclusive Dirichlet and Neumann subsets that are represented, respectively, by superscripts of lowercase letters $u$, $m$, and $g$ and those of uppercase letters $T$, $M$, and $G$, as $\Gamma_{\iota}=\Gamma_{\iota}^u\cup\Gamma_{\iota}^T=\Gamma_{\iota}^m\cup\Gamma_{\iota}^M$ and $\Upsilon_{\iota}=\Upsilon_{\iota}^g\cup\Upsilon_{\iota}^G$.  
We also denote by $\Gamma^u=\cup_{\iota}\Gamma_{\iota}^u$, $\Gamma^T=\cup_{\iota}\Gamma_{\iota}^T$, $\Gamma^m=\cup_{\iota}\Gamma_{\iota}^m$, $\Gamma^M=\cup_{\iota}\Gamma_{\iota}^M$,  $\Upsilon^g=\cup_{\iota}\Upsilon_{\iota}^g$, and $\Upsilon^G=\cup_{\iota}\Upsilon_{\iota}^G$ the unions of the Dirichlet and Neumann boundaries.  
As in \cite{Toupin1964}, coordinate derivatives of a scalar function $\phi$ are decomposed on $\Gamma$ into normal and tangential components as:
\begin{align}
\phi_{,J}=D\phi N_J+D_J\phi,\notag
\end{align}
where
\begin{align}
D\phi   &:=\phi_{,K}N_K,\notag\\
D_J\phi &:=\phi_{,J}-\phi_{,K}N_KN_J,\notag
\end{align}
where $N_J$ represents the components of the unit outward normal to $\Gamma$.  
Here as elsewhere ${(\hspace{1pt}\cdot\hspace{1pt})_{,J}}$ denotes the spatial derivative with respect to the reference coordinate variable $X_J$.  

Dirichlet boundary conditions for the displacement field $\boldsymbol{u}$ can now be given as:
\begin{align}
u_i=\bar{u}_i\quad\textrm{on }\Gamma^u,\quad
Du_i=\bar{m}_i\quad\textrm{on }\Gamma^m,\quad
u_i=\bar{g}_i\quad\textrm{on }\Upsilon^g,\label{E:dirichlet}
\end{align}
where $u_i$ $(i=1,2,3)$ are the components of $\boldsymbol{u}$ and $\bar{u}_i$, $\bar{m}_i$, and $\bar{g}_i$ are the components of known vector functions on $\Gamma^u$, $\Gamma^m$, and $\Upsilon^g$.  
On the other hand, we denote the components of the standard surface traction on $\Gamma^T$, the higher-order traction on $\Gamma^M$, and the line traction on $\Upsilon^G$ by $\bar{T}_i$, $\bar{M}_i$, and $\bar{G}_i$, whose mathematical formulas will be clarified shortly.  

\begin{figure}[]
    \begin{center}
    \begin{subfigure}[]{3in}
        \includegraphics[scale=0.5]{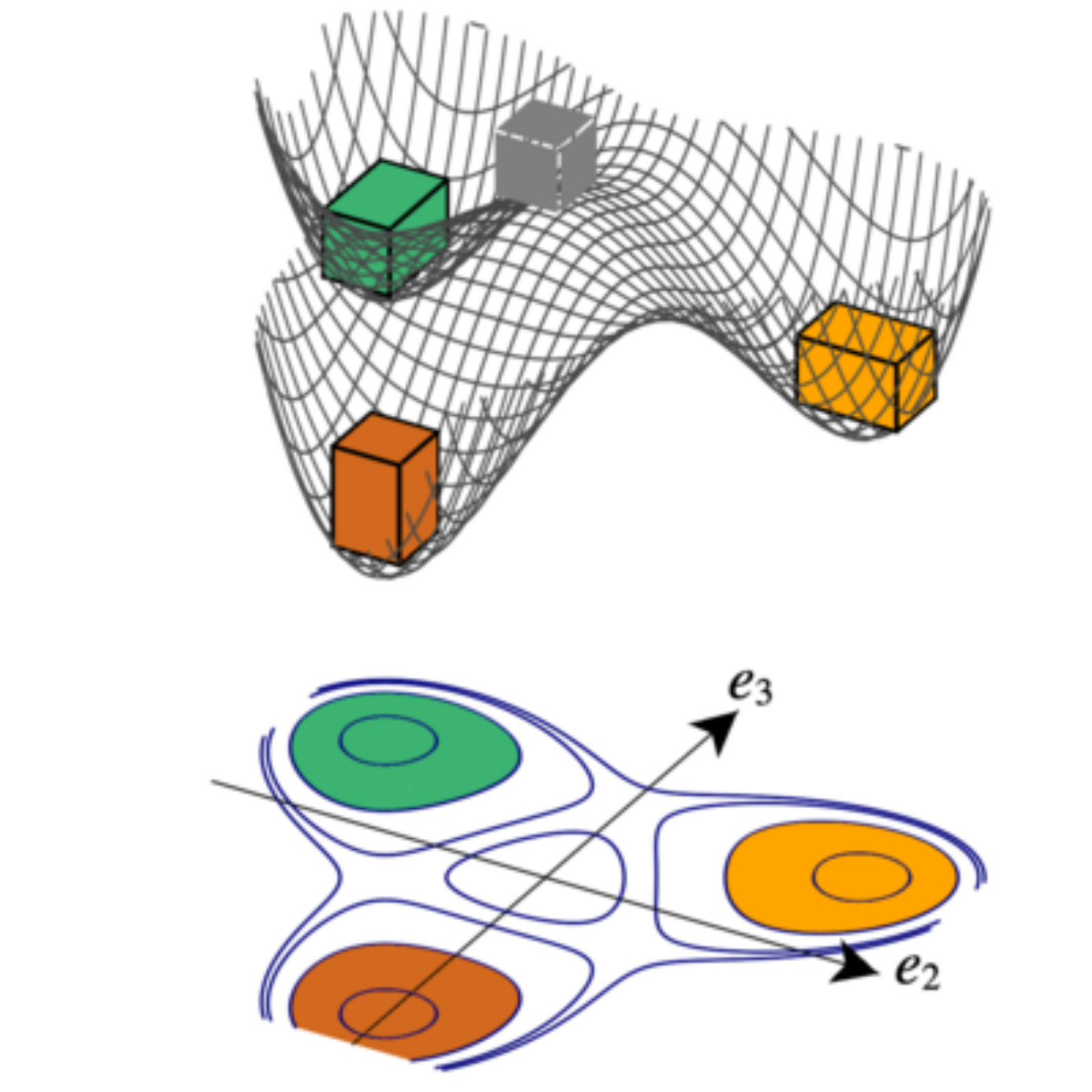}
        \caption{}
        \label{Fi:three-well}
    \end{subfigure}
    ~
    \begin{subfigure}[]{1.5in}
        \includegraphics[scale=0.15]{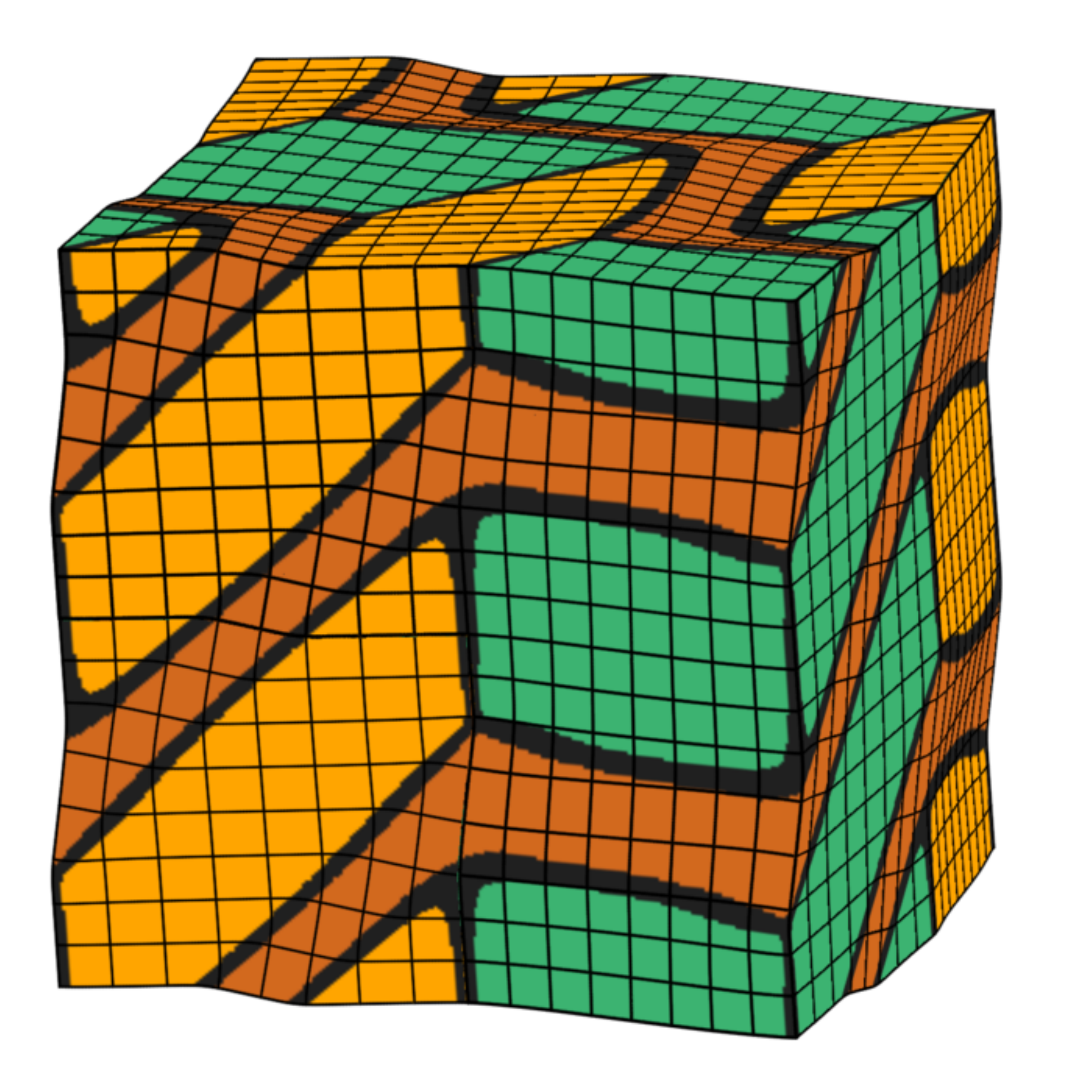}
        \caption{}
        \label{Fi:m0_111}
    \end{subfigure}
    ~
    \begin{subfigure}[]{1.5in}
        \includegraphics[scale=0.15]{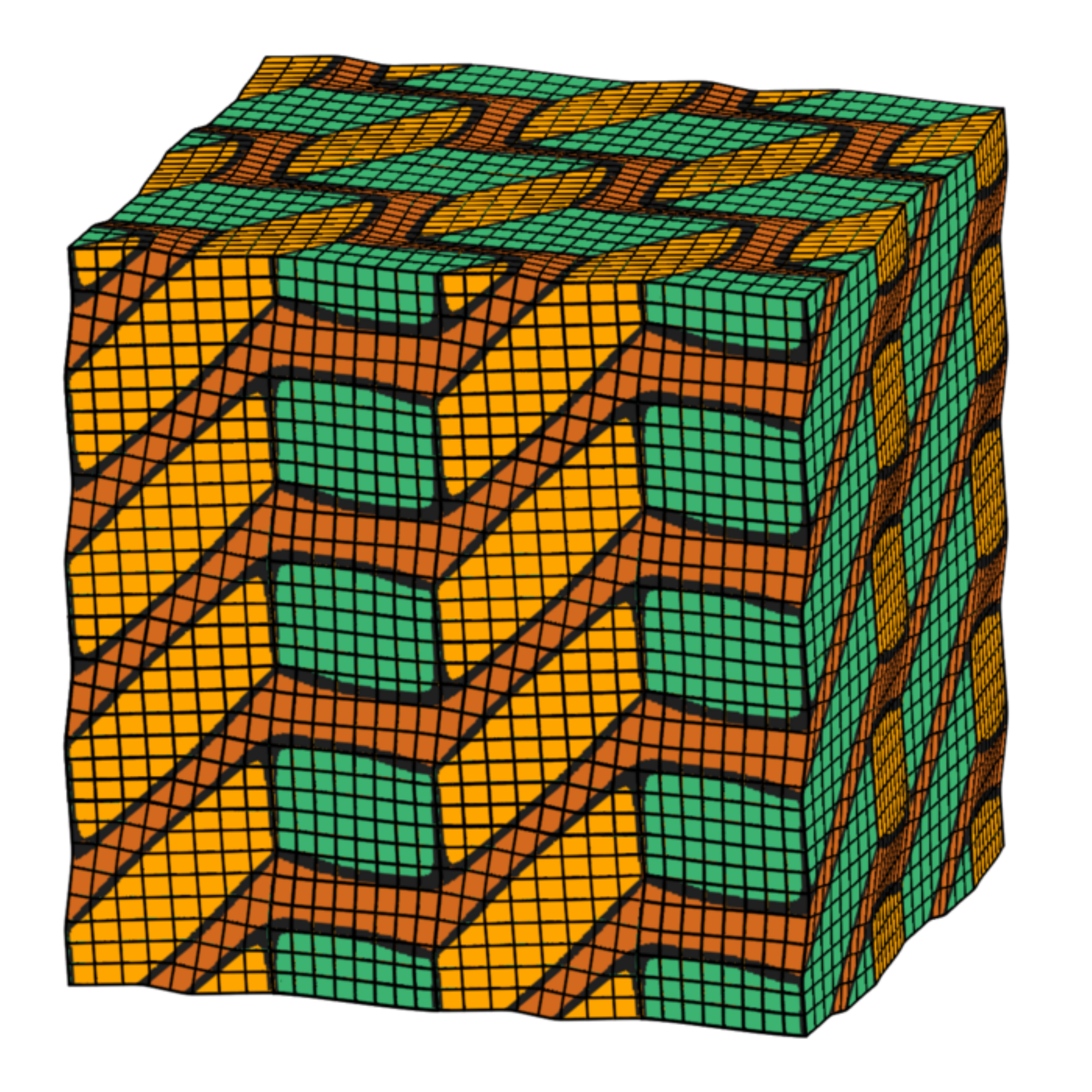}
        \caption{} 
        \label{Fi:m0_222}
    \end{subfigure}
    ~
    \caption{(\subref{Fi:three-well}) A surface plot on the $e_2-e_3$ space of the non-convex part of the free energy density function $\Psi$.  Energetically favored $X-$, $Y-$, and $Z-$oriented tetragonal variants are shown schematically at the bottom of the wells in \emph{orange}, \emph{green}, and \emph{brown}, respectively.  The energetically unfavored reference cubic variant is also shown at $(e_2,e_3)=(0,0)$.  The free energy density function is non-dimensionalized so that the wells have a unit depth. (\subref{Fi:m0_111}) The unit cell for the martensitic microstructure, obtained with the gradient elasticity model, and to be used for numerical homogenization. (\subref{Fi:m0_222}) The corresponding periodic martensitic microstructure. }
    \end{center}
\end{figure}

We derive the BVPs using a variational argument.  
The total free energy is a functional of $\boldsymbol{u}$ defined as:
\begin{align}
\Pi\left[\boldsymbol{u}\right]
:=\int_{\Omega}\Psi\mathrm{d} V
-\int_{\Gamma^T}u_i\bar{T}_i\mathrm{d} S
-\int_{\Gamma^M}Du_i\bar{M}_i\mathrm{d} S
-\int_{\Upsilon^G}u_i\bar{G}_i\mathrm{d} C,
\label{E:Pi}
\end{align}
where $\Psi=\tilde{\Psi}(F_{11},F_{12},\dots,F_{33},\dots,F_{11,1},F_{11,2},\dots,F_{33,3})$ is the non-dimensionalized free energy density function that is a function of the components of the deformation gradient tensor, $F_{iJ}=\delta_{iJ}+u_{i,J}$, and the gradient of the deformation gradient tensor, $F_{iJ,K}$, at each point $\boldsymbol{X}\in\Omega$.  
In the following, to facilitate formulation, we let $\boldsymbol{\zeta}$ be a short-hand notation of the array of all the components, $F_{11},F_{12},\dots,F_{33},\dots,F_{11,1},F_{11,2},\dots,F_{33,3}$, and write, e.g., $\tilde{\Psi}(F_{11},F_{12},\dots,F_{33},\dots,F_{11,1},F_{11,2},\dots,F_{33,3})$ as $\tilde{\Psi}(\boldsymbol{\zeta})$. 
This free-energy density function $\Psi$ that we consider in this work is defined as:
\begin{subequations}
\begin{align}
\Psi&:
=B_1e_1^2+B_2\left(e_2^2+e_3^2\right)+B_3e_3\left(e_3^2-3e_2^2\right)+B_4\left(e_2^2+e_3^2\right)^2+B_5\left(e_4^2+e_5^2+e_6^2\right)\notag\\
&\phantom{:}+l^2(e_{2,1}^2+e_{2,2}^2+e_{2,3}^2+e_{3,1}^2+e_{3,2}^2+e_{3,3}^2),
\label{E:Psi_e}
\end{align}
\label{E:Psi}%
\end{subequations}
where $B_1,...,B_5$ are constant with $B_1$, $B_4$, and $B_5$ positive, $l$ is the length scale parameter, and $e_1,...,e_6$ are reparameterized strains defined as:
\begin{subequations}
\begin{align}
e_1&=(E_{11}+E_{22}+E_{33})/\sqrt{3},\label{E:ea}\\
e_2&=(E_{11}-E_{22})/\sqrt{2},\\
e_3&=(E_{11}+E_{22}-2E_{33})/\sqrt{6},\\
e_4&=E_{23}=E_{32},\\
e_5&=E_{13}=E_{31},\\
e_6&=E_{12}=E_{21},\label{E:ef}
\end{align}
\label{E:e}%
\end{subequations}
where $E_{IJ}=1/2(F_{kI}F_{kJ}-\delta_{IJ})$ are the components of the Green-Lagrangian strain tensor.  
The free energy density \eqref{E:Psi} is non-convex with respect to the strain variables $e_2$ and $e_3$ with minima, or \emph{wells}, located to represent three energetically favored symmetric tetragonal variants and local maximum located to represent an energetically unfavored cubic variant; see Fig.\ref{Fi:three-well}.  
The parameters $B_1,...,B_5$ determine its landscape.  
Note that these pure tetragonal variants can be compatible with each other, but in general not with prescribed Dirichlet boundary conditions.  Arbitrarily fine layering of these tetragonal variants would mathematically resolve this incompatibility, but such microstructure would be non-realistic.  This non-physical behavior is prevented by the inclusion of strain-gradient terms in Eqn. \eqref{E:Psi}, which penalize rapid spatial changes of strain, or, equivalently, penalize arbitrarily large interface areas between different variants; strain-gradient terms in Eqn. \eqref{E:Psi} can thus be tied to an interfacial energy density.  The length scale parameter $l$ prescribes the level of compromise between fineness and incompatibility.


To formulate the BVPs, we take the variational derivative of the total free energy \eqref{E:Pi} with respect to $\boldsymbol{u}$ that satisfies the Dirichlet boundary conditions \eqref{E:dirichlet}.  
The test function $\boldsymbol{w}$ is then to satisfy:
\begin{align}
w_i=0\quad\textrm{on }\Gamma^u,\quad
Dw_i=0\quad\textrm{on }\Gamma^m,\quad
w_i=0\quad\textrm{on }\Upsilon^g,\label{E:admissible}
\end{align}
where $w_i$ are the components of $\boldsymbol{w}$.  
The variational derivative with respect to $\boldsymbol{u}$ is then obtained as:
\begin{align}
\delta_{\boldsymbol{u}}\Pi[\boldsymbol{u}]
&=\!\left.\frac{\mathrm{d}}{\mathrm{d}\varepsilon}\Pi[\boldsymbol{u}+\varepsilon\boldsymbol{w}]\right|_{\varepsilon=0}\notag\\
&=\int_{\Omega}\left(w_{i,J}P_{iJ}+w_{i,JK}B_{iJK}\right)\mathrm{d} V
-\int_{\Gamma^T}w_i\bar{T}_i\mathrm{d} S
-\int_{\Gamma^M}Dw_i\bar{M}_i\mathrm{d} S
-\int_{\Upsilon^G}w_i\bar{G}_i\mathrm{d} C,\label{E:bvp_weak_temp}
\end{align}
where $P_{iJ}=\tilde{P}_{iJ}(\boldsymbol{\zeta})$ are the components of the first Piola-Kirchhoff stress tensor and $B_{iJK}=\tilde{B}_{iJK}(\boldsymbol{\zeta})$ are the components of the higher-order stress tensor that are defined as:
\begin{align}
\tilde{P}_{iJ}&:=\frac{\partial\tilde{\Psi}}{\partial F_{iJ}},\notag\\
\tilde{B}_{iJK}&:=\frac{\partial\tilde{\Psi}}{\partial F_{iJ,K}}.\notag
\end{align}
At equilibrium one has $\delta_{\boldsymbol{u}}\Pi[\boldsymbol{u}]=0$.  We then have from \eqref{E:bvp_weak_temp}:
\begin{align}
\int_{\Omega}\left(w_{i,J}\tilde{P}_{iJ}(\boldsymbol{\zeta})+w_{i,JK}\tilde{B}_{iJK}(\boldsymbol{\zeta})\right)\mathrm{d} V
-\int_{\Gamma^T}w_i\bar{T}_i\mathrm{d} S
-\int_{\Gamma^M}Dw_i\bar{M}_i\mathrm{d} S
-\int_{\Upsilon^G}w_i\bar{G}_i\mathrm{d} C=0.\label{E:bvp_weak}
\end{align}
Eqns. \eqref{E:bvp_weak}, \eqref{E:dirichlet}, and \eqref{E:admissible} define the weak form of the BVPs.  

The variational argument can further lead us to identify the strong form and the Neumann boundary conditions corresponding to \eqref{E:bvp_weak} as the following:
\begin{subequations}
\begin{align}
-P_{iJ,J}+B_{iJK,JK}&=0\hfill                                                                &&\text{in }\Omega,\\
P_{iJ}N_J-B_{iJK,K}N_J-D_J(B_{iJK}N_K)+B_{iJK}\left(b_{LL} N_JN_K-b_{JK}\right)&=\bar{T}_i   &&\textrm{on }\Gamma^T,\\
B_{iJK}N_KN_J&=\bar{M}_i                                                                     &&\textrm{on }\Gamma^M,\\
[\![B_{iJK}N_KN_J^\Gamma]\!]&=\bar{G}_i                                                      &&\textrm{on }\Upsilon^G,
\end{align}
\label{E:bvp_strong}%
\end{subequations}
where $b_{IJ}$ are the components of the second fundamental form on $\Gamma^T$, $N^{\Gamma}_J$ are the components of the unit outward normal to the boundary curve $\Upsilon_{\iota}\subset\overline{\Gamma_{\iota'}}$, and, on each $\Upsilon^G_\iota$, $[\![B_{iJK}N_KN_J^\Gamma]\!]:=B_{iJK}N^+_KN_J^{\Gamma+}+B_{iJK}N^-_KN_J^{\Gamma-}$ is the \emph{jump}, where superscripts $+$ and $-$ represent two surfaces sharing $\Upsilon^G_\iota$; see \cite{Toupin1964} for details.

Equations \eqref{E:dirichlet} and \eqref{E:admissible}-\eqref{E:bvp_strong} describe the general BVP of gradient-coercified non-convex hyperelasticity at finite strain. As in Sec~\ref{sec:homogmeth} we use periodic boundary conditions on $\Omega = (0,1)^3$ for DNS data generation.

All computations were carried out using isogeometric analysis (IGA) \cite{Cottrell2009} within the mechanoChem library available at \url{https://github.com/mechanoChem/mechanoChem}. The computational framework for the above model of gradient-coercified non-convex hyperelasticity at finite strain has been described elsewhere \cite{Rudraraju2014,Sagiyamaetal2016,SagiyamaGarikipati2017b}.

\subsection{Macroscopic modeling for a single microstructure}\label{SS:ge_macro_single}
Macroscopic quantities of interest are the macroscopic Green-Lagrange strain $\overline{\boldsymbol{E}}$, and the corresponding macroscopic free energy density function $\overline{\Psi}$ and macroscopic second Piola-Kirchhoff stress $\overline{\boldsymbol{S}}$.  
The macroscopic Green-Lagrange strain is defined as:
\begin{align}
    \overline{\boldsymbol{E}}=\frac{1}{2}(\overline{\boldsymbol{F}}^\textrm{T}\overline{\boldsymbol{F}}-\boldsymbol{I}),
\end{align}
and the macroscopic free energy density function is defined as:
\begin{align}
    \overline{\Psi}=\int\limits_\Omega \Psi\mathrm{d} V.
\end{align}
The macroscopic second Piola-Kirchhoff stress is computed as:
\begin{align}
    \overline{\boldsymbol{S}}=\overline{\boldsymbol{F}}^{-1}\overline{\boldsymbol{P}},
\end{align}
where $\overline{\boldsymbol{P}}$ is the macroscopic first Piola-Kirchhoff stress tensor, which in turn is computed from the surface average of the effective boundary traction represented by the boundary integral terms in the weak form \eqref{E:bvp_weak}.
In this section we aim to discover a hyperelastic constitutive relation of the form \eqref{eq:DNN_hyperelas} for this homogenized material.  
In contrast to the synthetic example presented in Sec.~\ref{sec:homogmeth}, we have no previous knowledge on the nature of this homogenized material, which makes this observation rather meaningful.  

\subsubsection{Data sampling}\label{SSS:ge_macro_single_data}
We sampled $\overline{\boldsymbol{E}}$ in the six-dimensional subspace of the strain components, $[-0.1,0.1]^6$, and solved the BVP in weak form
\begin{align}
\int_{\Omega}\left(w_{i,J}\tilde{P}_{iJ}(\boldsymbol{\zeta})+w_{i,JK}\tilde{B}_{iJK}(\boldsymbol{\zeta})\right)\mathrm{d} V
=0.\label{E:pbvp_weak}
\end{align}
for each instance of average deformation gradient $\overline{\boldsymbol{F}}$ with periodic boundary conditions.

\begin{figure}
\begin{center}
    ~
        \includegraphics[scale=0.35]{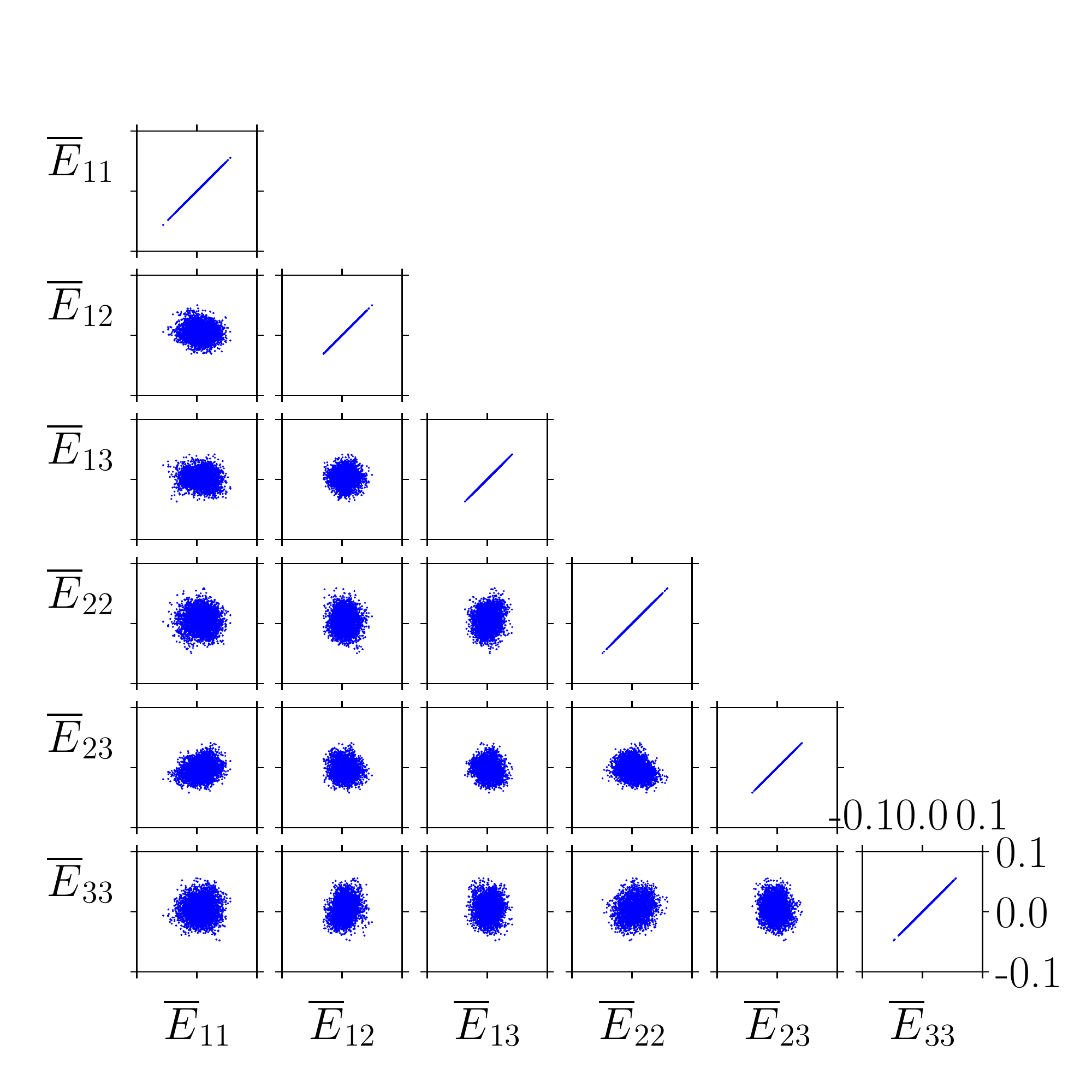}
\end{center}
\caption{Distribution of sample data points based on a Sobol sequence in $\mathbb{R}^6$: 2770 points were generated in $[-0.1,0.1]^6$ for gradient-coercified non-convex hyperelasticity at finite strain.  Projections of these points onto the two-dimensional hyperplanes $E_{IJ}-E_{I'J'}$ are shown.  }
\label{Fi:ge_dens}
\end{figure}

We then computed $\overline{\Psi}$ and $\overline{\boldsymbol{S}}$ for each solution in the postprocessing.  
To solve the BVP for a given $\overline{\boldsymbol{E}}$, we need a good initial guess to the solution.  
We consider a large Sobol sequence, $\{a_k\}$ $(k=1,\cdots,2^{24})$, in $[-0.1,0.1]^6$ and subsequently compute solutions at elements of $\{a_k\}$ as described below.  
We initially have a solution at $a_0=a_{k_0}$ corresponding to $\overline{\boldsymbol{F}}=\boldsymbol{I}$.  
Provided that we have computed solutions at $l-1$ points, $\{a_{k_0},\cdots,a_{k_{l-1}}\}$, the $l-$th solution is obtained as described below.
We first randomly choose an element $a_{k_l}'$ in $\{a_{k_0},\cdots,a_{a_{l-1}}\}$.  
We then randomly choose a point, at which solution has not yet computed, in a neighborhood of $a_{k_l}'$, and then compute the solution at $a_{k_l}$ using the solution at $a_{k_l}'$ as the initial guess.  
Here, we define a small neighborhood as a ball of radius $0.02$ centered at $a_{k_l}'$ in $[-0.1,0.1]^6$.  
We repeat this process and obtained 2,770 solutions, which produced data sets: $(\overline{\boldsymbol{E}}^{(i)}; \overline{\Psi}^{(i)}, \overline{\boldsymbol{S}}^{(i)})$ ($i=1,\cdots,2,770$).
Fig.\ref{Fi:ge_dens} shows the distribution of these data points in the strain component space $[-0.1,0.1]^6$.  
Although a slight bias is present, we regarded this as a good representation of the strain space.

\subsubsection{NN representation}\label{SSS:ge_macro_single_NN}
We then set up neural networks to express $\overline{\Psi}$ in terms of the six components of $\overline{\boldsymbol{E}}$, $\overline{E}_{IJ}$ ($I,J=1,2,3$).  
We used an open source machine learning framework TensorFlow \cite{tensorflow} for our problems.
We adopted fully connected neural networks of six inputs and one output.  
Various combinations of number of hidden layers and number of nodes per layer were tested, and the best combination was selected upon cross-validation analysis.  
In this work number of nodes per layer was fixed across all hidden layers.  
Rectified linear (ReLU) activation function was used on the hidden nodes and the linear function was used on the output node.

The 2,770 data sets were split into subsets of 1,792, 256, and 722 data sets for training, cross-validation, and testing.  
In training, for each given number of hidden layers and number of nodes per layer, training data set was used to optimize the model parameters, \emph{weights} and \emph{biases}.  
We used mean squared error (MSE) for the loss function and the Adam optimizer with learning rate $5\times 10^{-4}$ for the optimizer.  
In an attempt to minimize the generalized error we further split the training data set into 7 subsets, each containing 256 data points, and conduct k-fold cross-validation analysis with $k=7$.
\emph{Early stopping} was applied for each of the seven training phases as soon as the validation error starts increasing.  
Thus, for each combination of hyper parameters, we obtain seven trained networks.  
These networks are averaged to produce a representative neural network for the given hyper parameters.

\subsubsection{Numerical homogenization via a hyperelastic neural network representation}
\begin{figure}
\begin{center}
        \begin{subfigure}[b]{3in}
            \centering
            \includegraphics[scale=0.35]{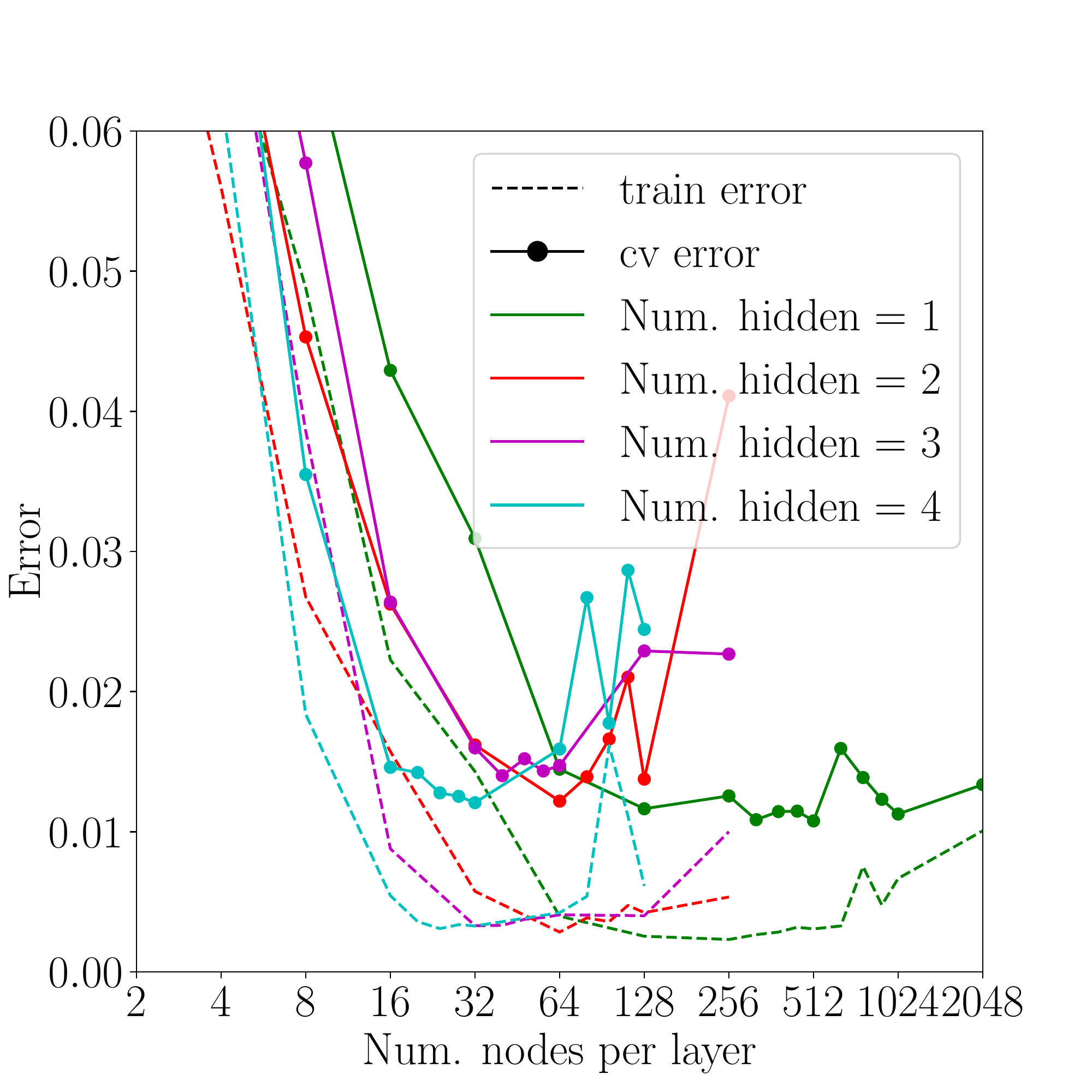}
            \caption{}
            \label{Fi:ge_cv_32}
        \end{subfigure}
        ~
        \begin{subfigure}[b]{3in}
            \centering
            \includegraphics[scale=0.35]{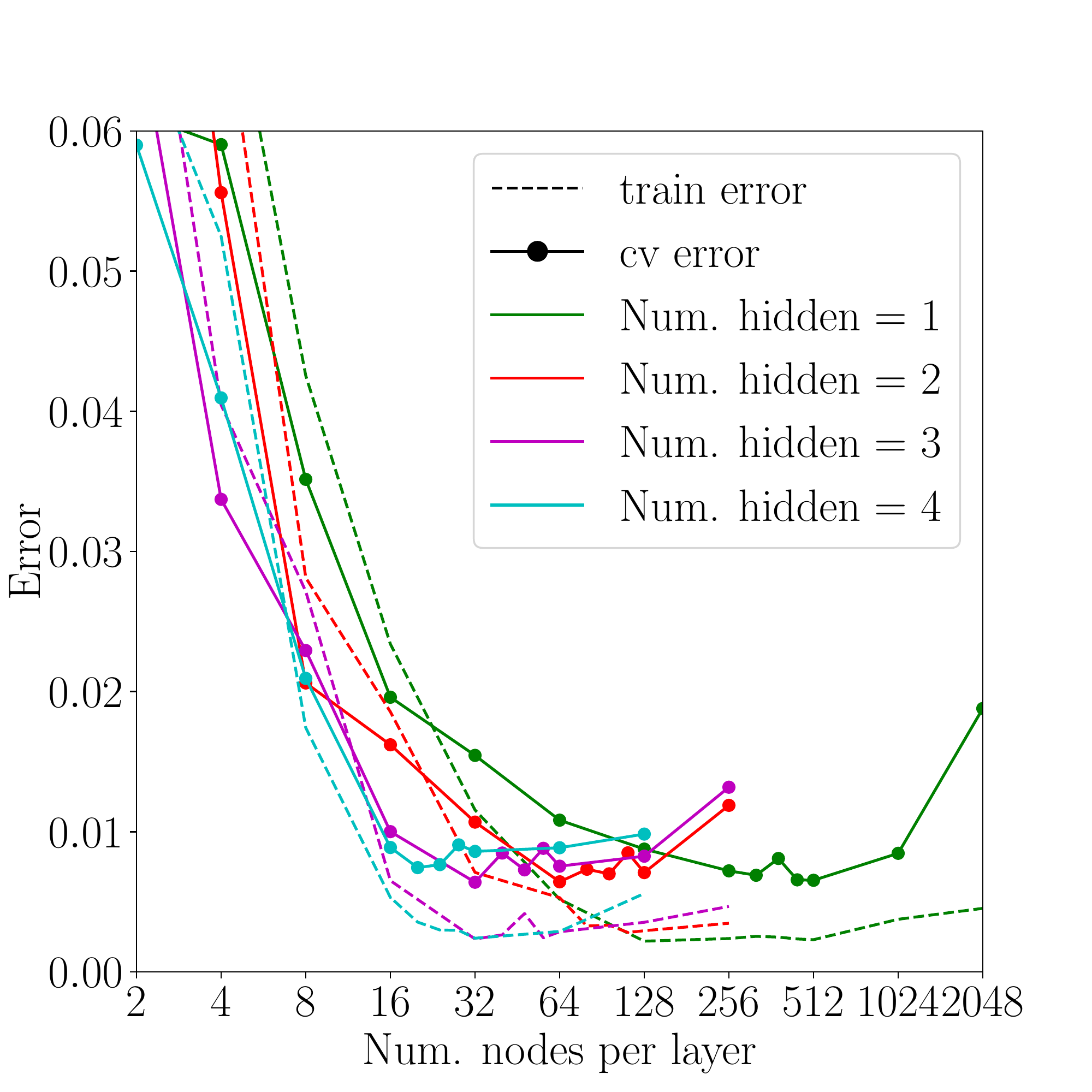}
            \caption{}
            \label{Fi:ge_cv_64}
        \end{subfigure}
        
        \begin{subfigure}[b]{3in}
            \centering
            \includegraphics[scale=0.35]{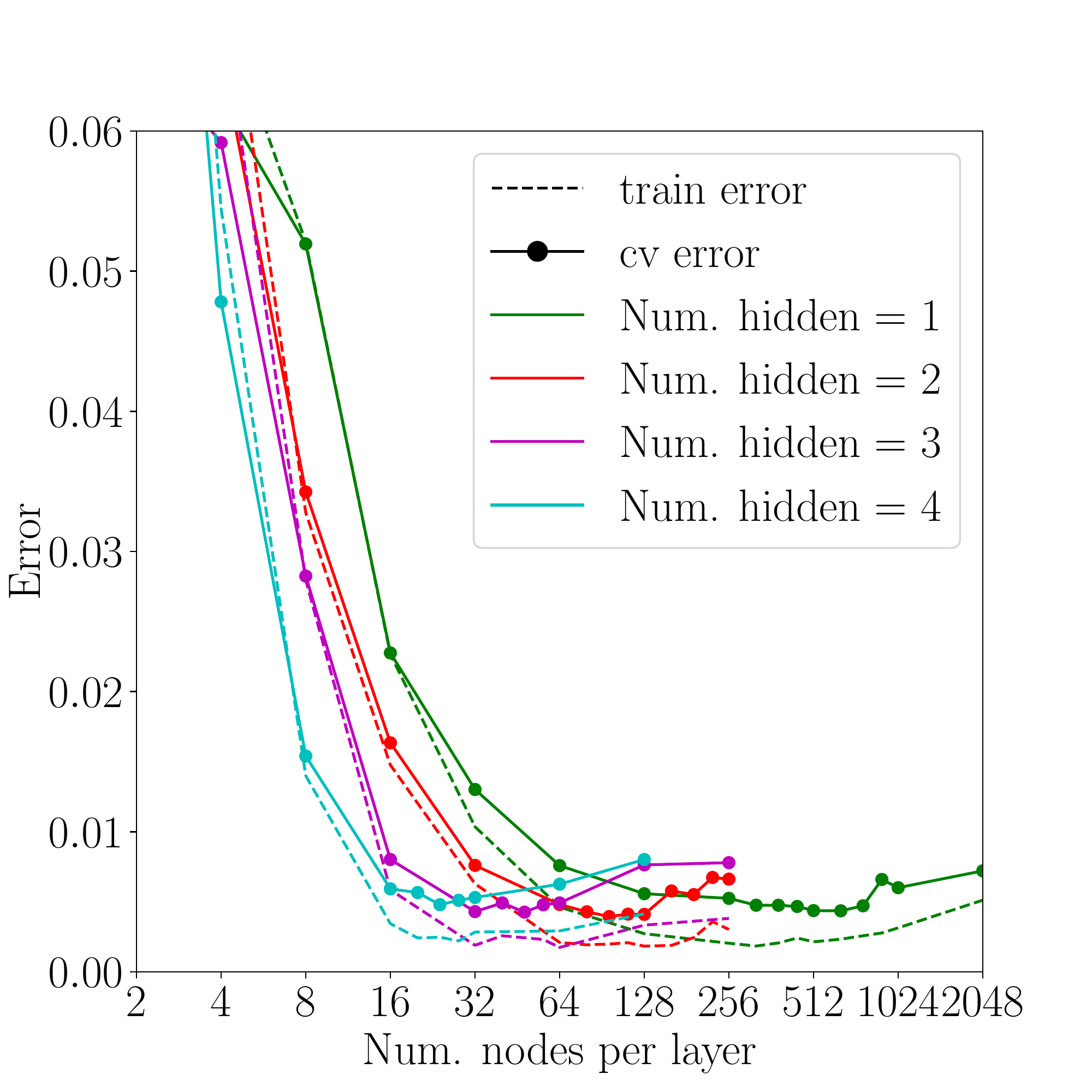}
            \caption{}
            \label{Fi:ge_cv_128}
        \end{subfigure}
        ~
        \begin{subfigure}[b]{3in}
            \centering
            \includegraphics[scale=0.35]{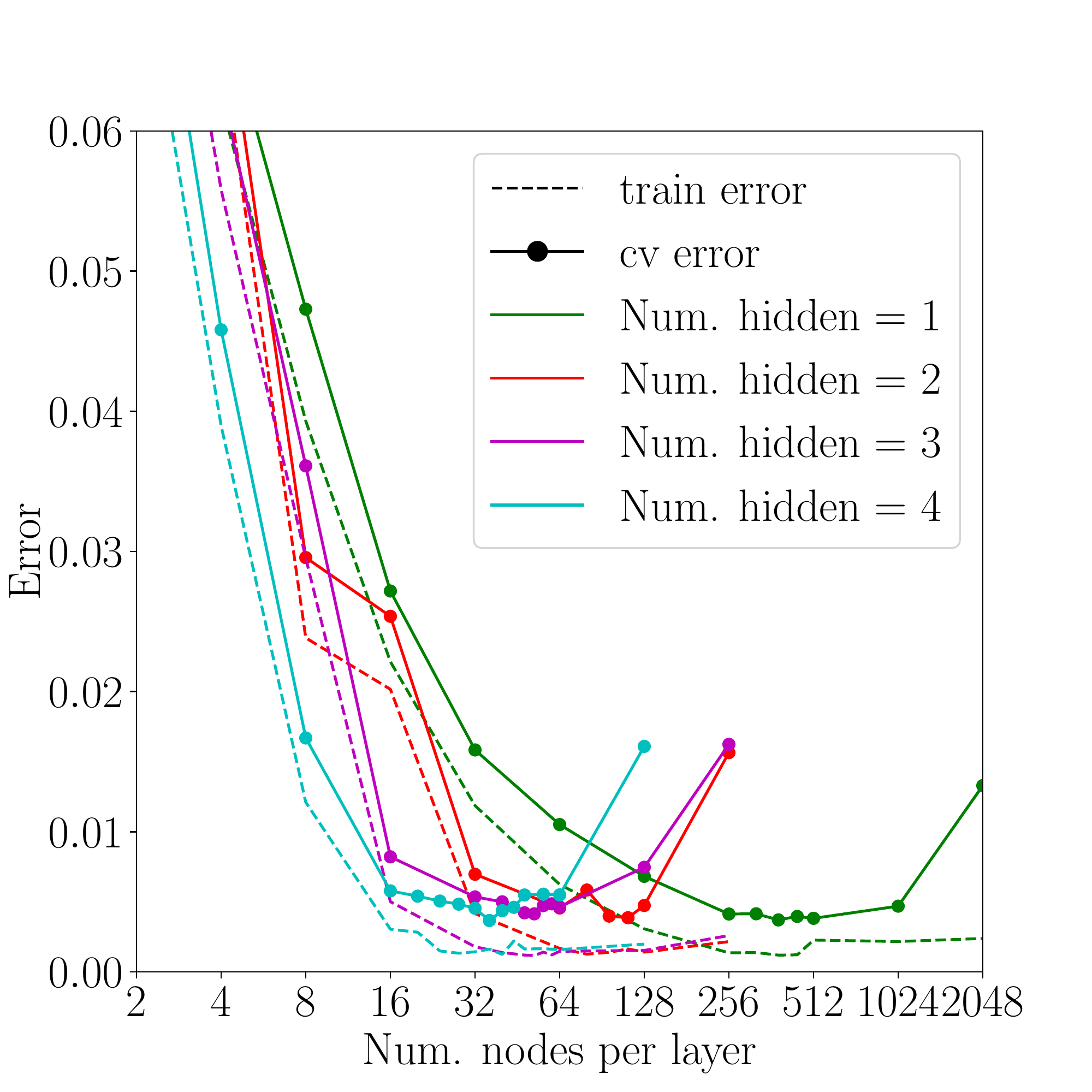}
            \caption{}
            \label{Fi:ge_cv_256}
        \end{subfigure}
\caption{ Training and cross-validation errors for (\subref{Fi:ge_cv_32}) $N=256$, (\subref{Fi:ge_cv_64}) $N=512$, (\subref{Fi:ge_cv_128}) $N=1024$, and (\subref{Fi:ge_cv_256}) $N=2048$ for data obtained with the martensitic microstructures arrived at using the gradient-coercified non-convex hyperelasticity at finite strain.  Errors are plotted in log-scale for different number of hidden layers, $H=1,2,3,4$, and various number of nodes per hidden layer, $O$, between 2 and 2048. }
\label{Fi:ge_cv}
\end{center}

\end{figure}

These trained neural networks for various combinations of hyper parameters are then tested against the cross-validation data set of 256 data points that were not used at all in training.  
The cross-validation errors are plotted in Fig. \ref{Fi:ge_cv} along with the training errors; square root of MSE was used as the measure of the error.
This figure implies that the neural network of one hidden layer with 384 nodes per layer is optimum, and produces a cross-validation error of 0.00371.  
This network was finally tested against the test data set of 722 data points and gave an error of 0.00337.  
These results are summarized in Table \ref{Ta:ge_cv}. The optimal neural network was then used to compute our prediction of the second Piola-Kirchhoff stress as:
\begin{align}
    \overline{\boldsymbol{S}}_\textrm{NN}=\frac{\partial\overline{\Psi}_\textrm{NN}}{\partial\overline{\boldsymbol{E}}},
\end{align}
This yields a hyperelastic NN representation for numerical homogenization of the microscopic model that produces the martensitic microstructures.

\begin{figure}
\begin{center}
        \begin{subfigure}[b]{3in}
            \centering
            \includegraphics[scale=0.35]{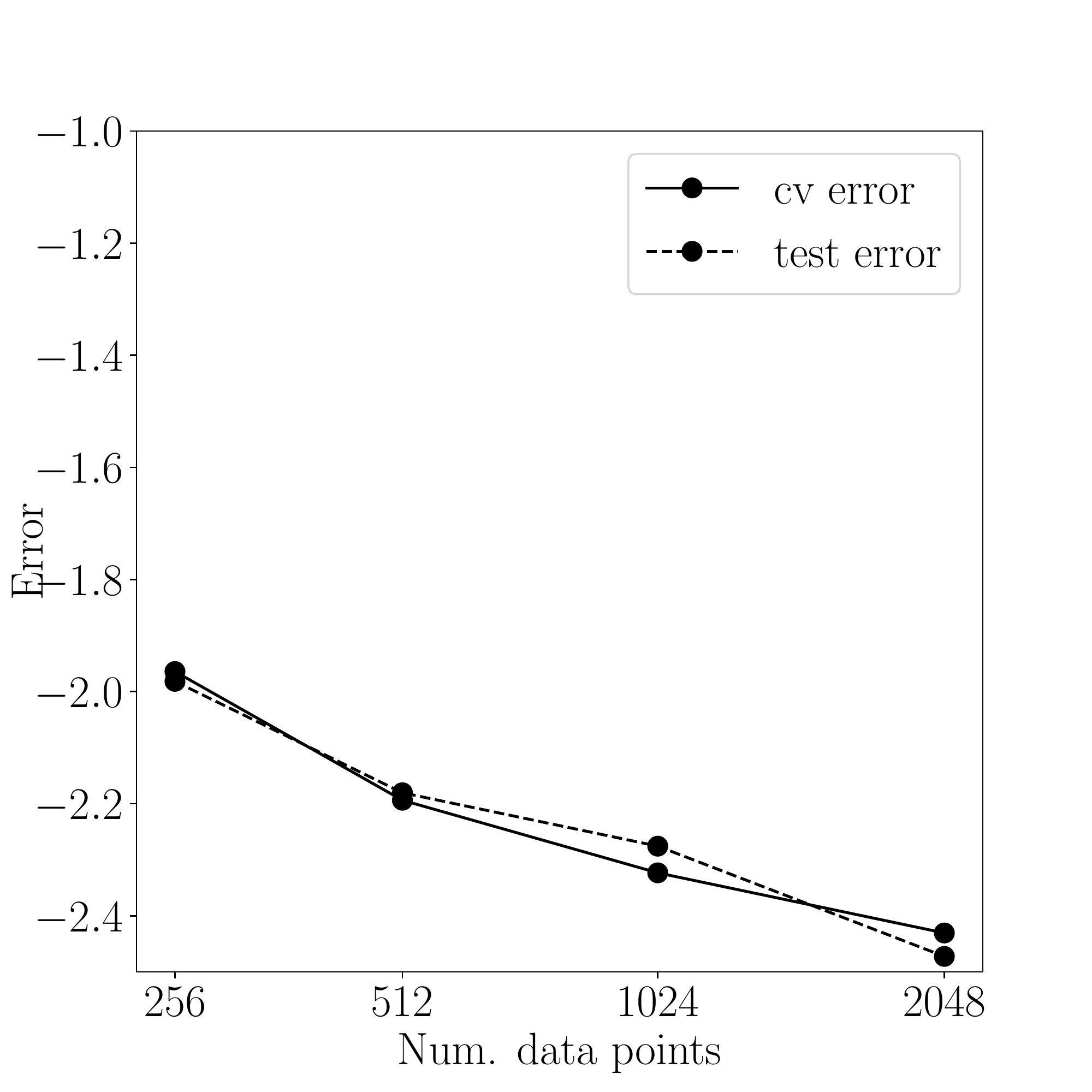}
            \caption{}
            \label{Fi:Uconv_ge}
        \end{subfigure}
         ~
       \begin{subfigure}[b]{3in}
            \centering
	    \includegraphics[scale=0.35]{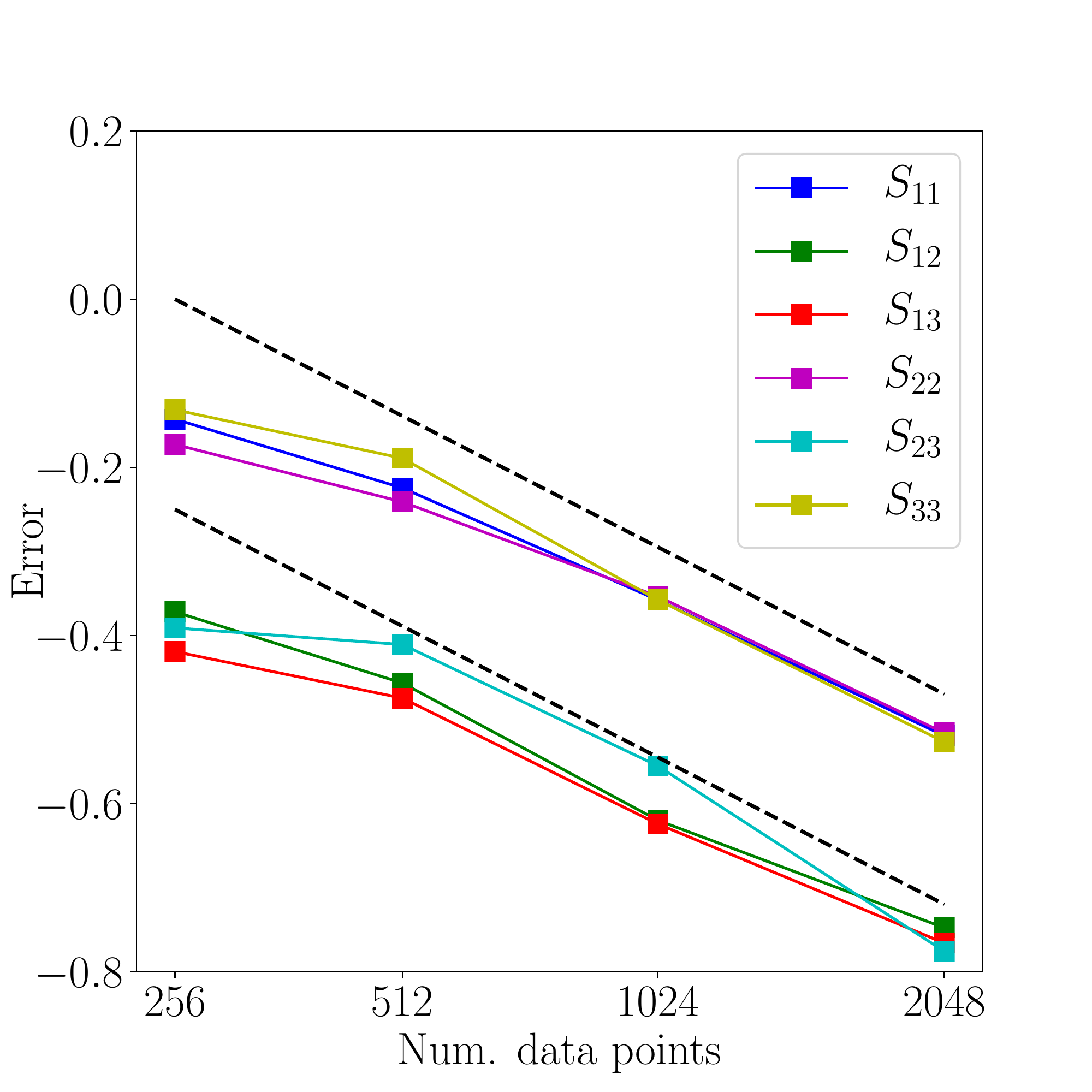}
            \caption{}
            \label{Fi:Sconv_ge}
        \end{subfigure}        
\caption{Convergence of neural network predictions to the \emph{correct} DNS data with respect to data resolution for (\subref{Fi:Uconv_ge}) $\overline{\Psi}_\mathrm{NN}$ and (\subref{Fi:Sconv_ge}) $\overline{\boldsymbol{S}}_\mathrm{NN}$ with gradient-coercified non-convex hyperelasticity at finite strain .  Cross-validation and test errors were separately computed for $\overline{\Psi}_\mathrm{NN}$ and all data were used for $\overline{\boldsymbol{S}}_\mathrm{NN}$.}
\label{fig:Conv_ge}
\end{center}
\end{figure}

\begin{table}
  \begin{tabular}{|c|c|c|c|c|}
    \hline
    N & H & O & cv & test\\
    \hline
    256 & 1 & 320 & 0.01086 & 0.01043\\
    \hline
    512 & 3 & 32 & 0.00640 & 0.00666\\
    \hline
    1024 & 1 & 384 & 0.00475 & 0.00530\\
    \hline
    2048 & 1 & 384 & 0.00371 & 0.00337\\
    \hline
  \end{tabular}
  \caption{Optimal number of hidden layers $H$ and number of nodes per hidden layer $O$ for $N=256,512,1024,2048$ from Fig. \ref{Fi:ge_cv} for gradient-coercified non-convex hyperelasticity at finite strain;  corresponding cross-validation and test errors are also shown.}%
  \label{Ta:ge_cv}
\end{table}

\begin{figure}
\begin{center}
        \begin{subfigure}[b]{3in}
            \centering
            \includegraphics[scale=0.35]{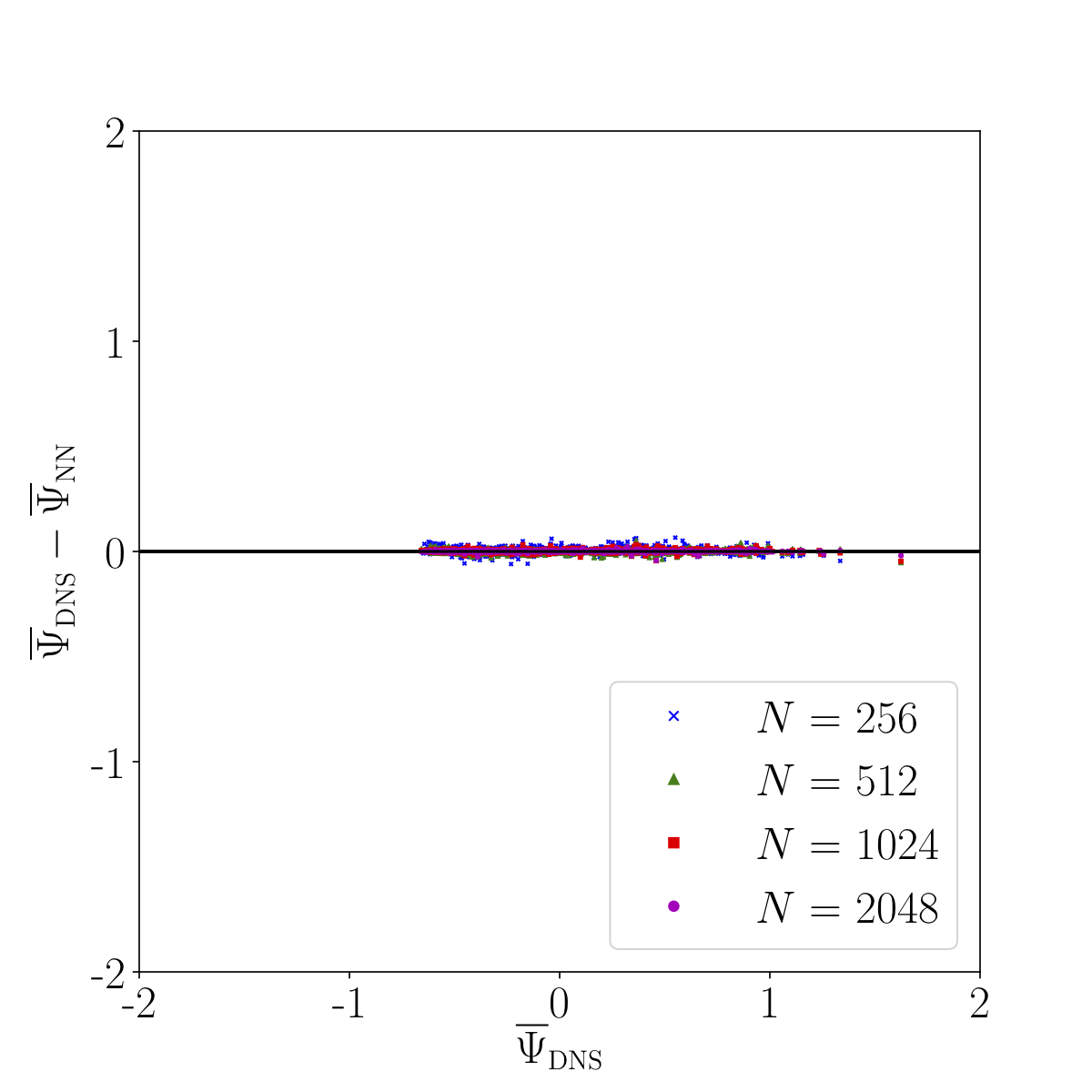}
            \caption{}
            \label{Fi:ge_Uerror}
        \end{subfigure}
        ~
        \begin{subfigure}[b]{3in}
            \centering
            \includegraphics[scale=0.35]{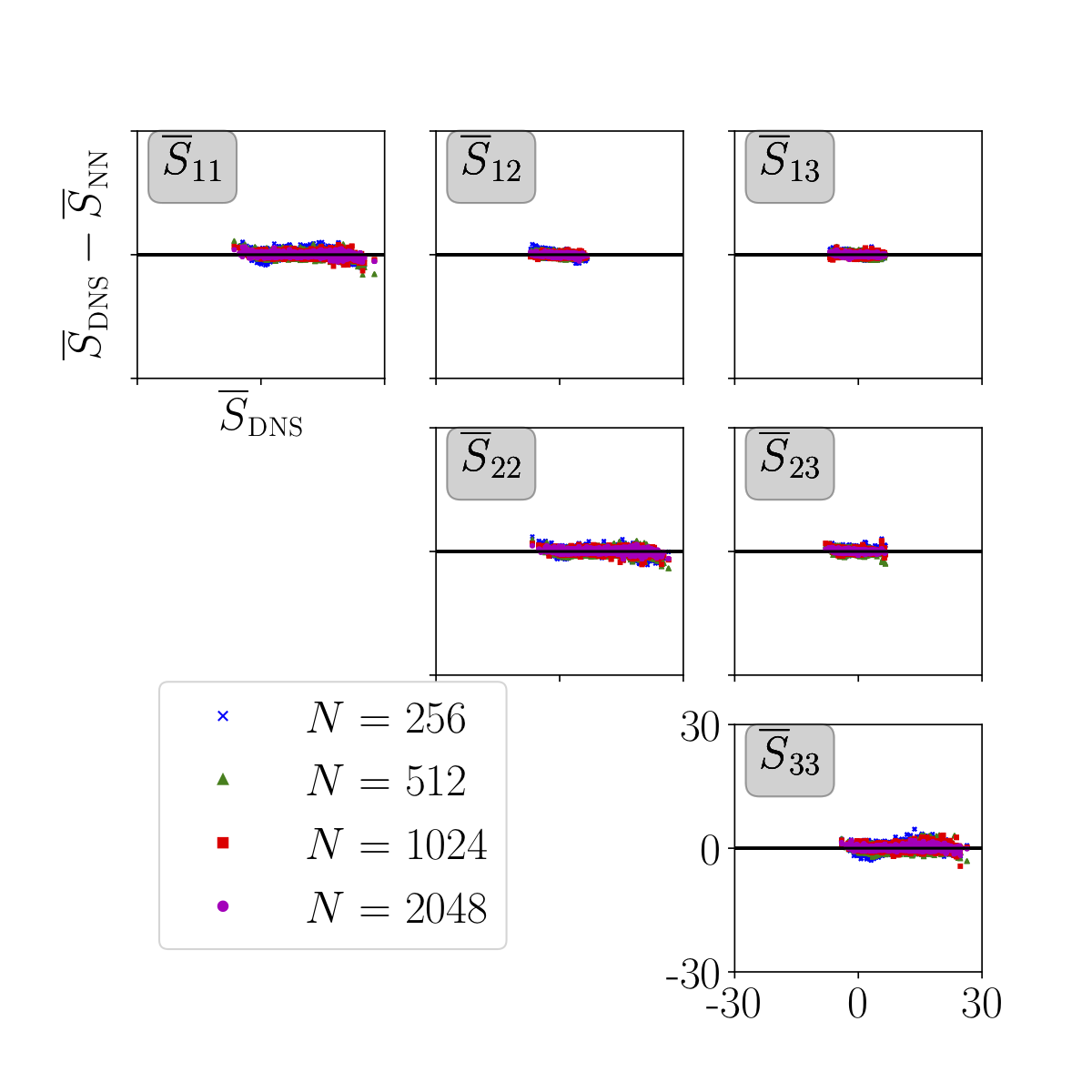}
            \caption{}
            \label{Fi:ge_Serror}
        \end{subfigure}
\caption{  Absolute error of the neural network predictions when compared to the \emph{correct} values from DNS for (\subref{Fi:ge_Uerror}) $\overline{\Psi}$ and (\subref{Fi:ge_Serror}) $\overline{S}_{IJ}$ with gradient-coercified non-convex hyperelasticity at finite strain.  Neural network predictions are computed and are compared to DNS data for all 2770 data points using the optimal network structures obtained in Table \ref{Ta:ge_cv} for each $N$ . }
\end{center}
\end{figure}

We introduce the following measures of error for the free energy and the second-Piola stress components:
\begin{subequations}
\begin{align}
    \overline{\psi}&=\overline{\Psi}_\textrm{NN}-\overline{\Psi}_\textrm{DNS},\\
    \overline{\boldsymbol{S}}&=\overline{\boldsymbol{S}}_\textrm{NN}-\overline{\boldsymbol{S}}_\textrm{DNS},
\end{align}
\end{subequations}

To further validate our approach, we studied convergence of the error with respect to the data set size.  
Specifically, we subsampled $256,512,$ and $1024$ data points from the original 2048 training data points, and repeated the same analysis as above with $N_\textrm{train}=256,512,1024$.  Fig.~\ref{Fi:Uconv_ge} plots $\log_{10}\overline{\Psi}$ against $N^{1/6}$ and Fig.~\ref{Fi:Sconv_ge} plots $\log_{10}\overline{S}_{IJ}$ against $N^{1/6}$.  Convergence rates for all stress components are virtually the same. The dashed lines confirm that the stress components all converge as $\sim N^{-1/6}$, as was the case with the neo-Hookean model in Fig. \ref{Fi:nh_Sconv}. This comparison shows that the DNN representation yields a viable, hyperelastic, homogenized constitutive law for martensitic microstructures formed from a microscopic model of gradient coercified non-convex hyperelasticity at finite strain. Furthermore this machine learned homogenization retains the same characteristics of convergence, and hence fidelity as we found with the similar representation of the classical, neo-Hookean model.

In Figure \ref{Fi:ge_Uerror} we compare the absolute error in $\overline{\Psi}_\textrm{NN}$ relative to $\overline{\Psi}_\textrm{DNS}$. For each data point, the  optimal hyper parameters from Table \ref{Ta:ge_cv} were used. Similarly, $\overline{\boldsymbol{\boldsymbol{S}}}_\textrm{NN}$ was compared to $\overline{\boldsymbol{S}}_\textrm{DNS}$.  
Fig. \ref{Fi:ge_Serror} shows this comparison componentwise.  We observe that the absolute errors are low for $\overline{\Psi}_\textrm{NN}$, as well as for $ \overline{\boldsymbol{S}}_\textrm{NN}$ when compared against the corresponding errors obtained for the DNN representation of the neo-Hookean model in Figs. \ref{Fi:nh_Uerror} and \ref{Fi:nh_Serror}. However, this is likely due to the sparseness of data at higher energies and stresses---thus the data did not probe the deep nonlinear regime.

\section{Conclusion}\label{sec:conclusion}

We have presented an approach to numerical homogenization of the hyperelastic response of martensitic microstructures using DNN representations. Our studies comes with detailed optimization over hyper parameters, and learning curves There are several important highlights to our results;

\begin{figure}
    \begin{subfigure}[b]{1.5in}
        \centering
        \includegraphics[scale=0.15]{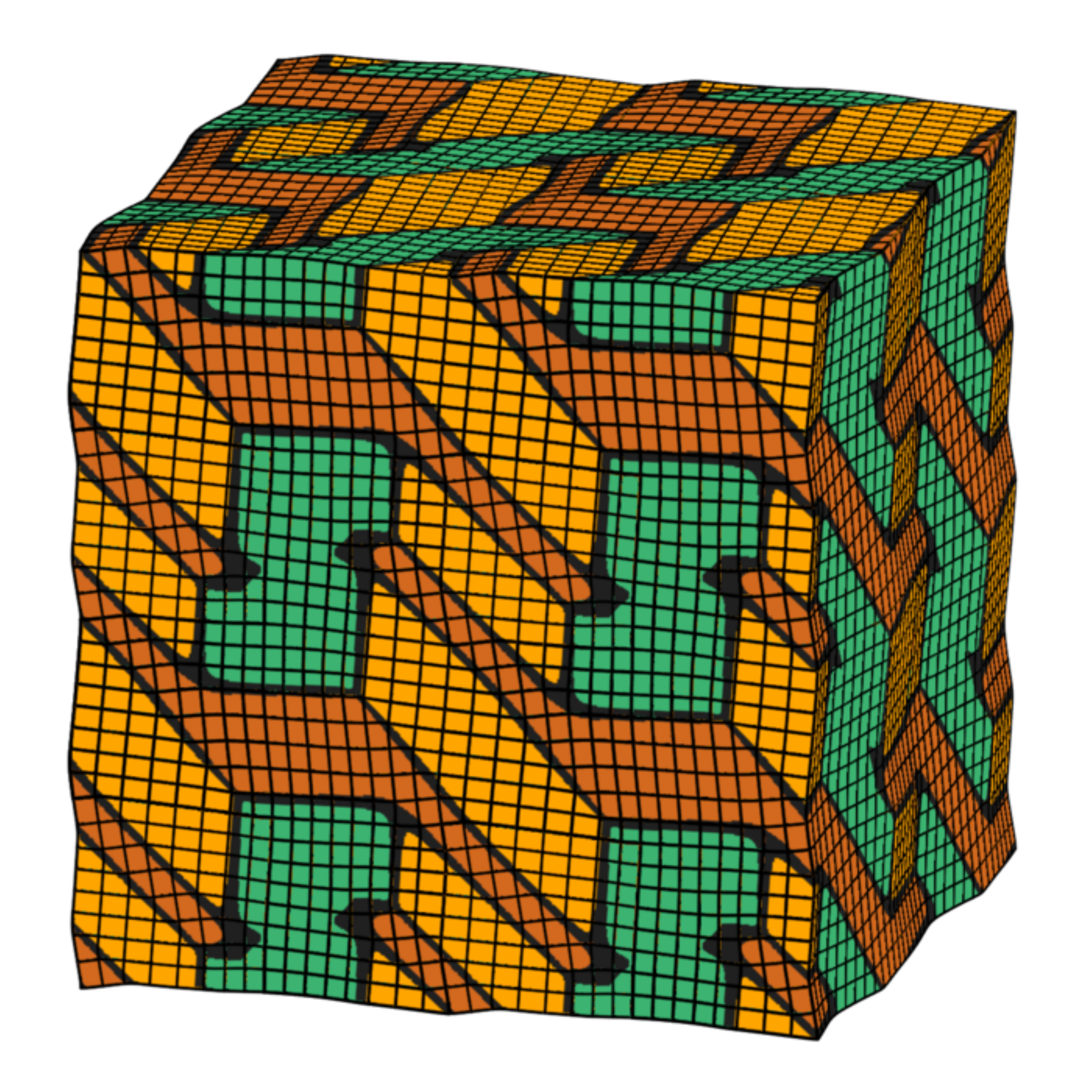}
        \caption{}
        \label{Fi:m7_222}
    \end{subfigure}
    ~
    \begin{subfigure}[b]{1.5in}
        \centering
        \includegraphics[scale=0.15]{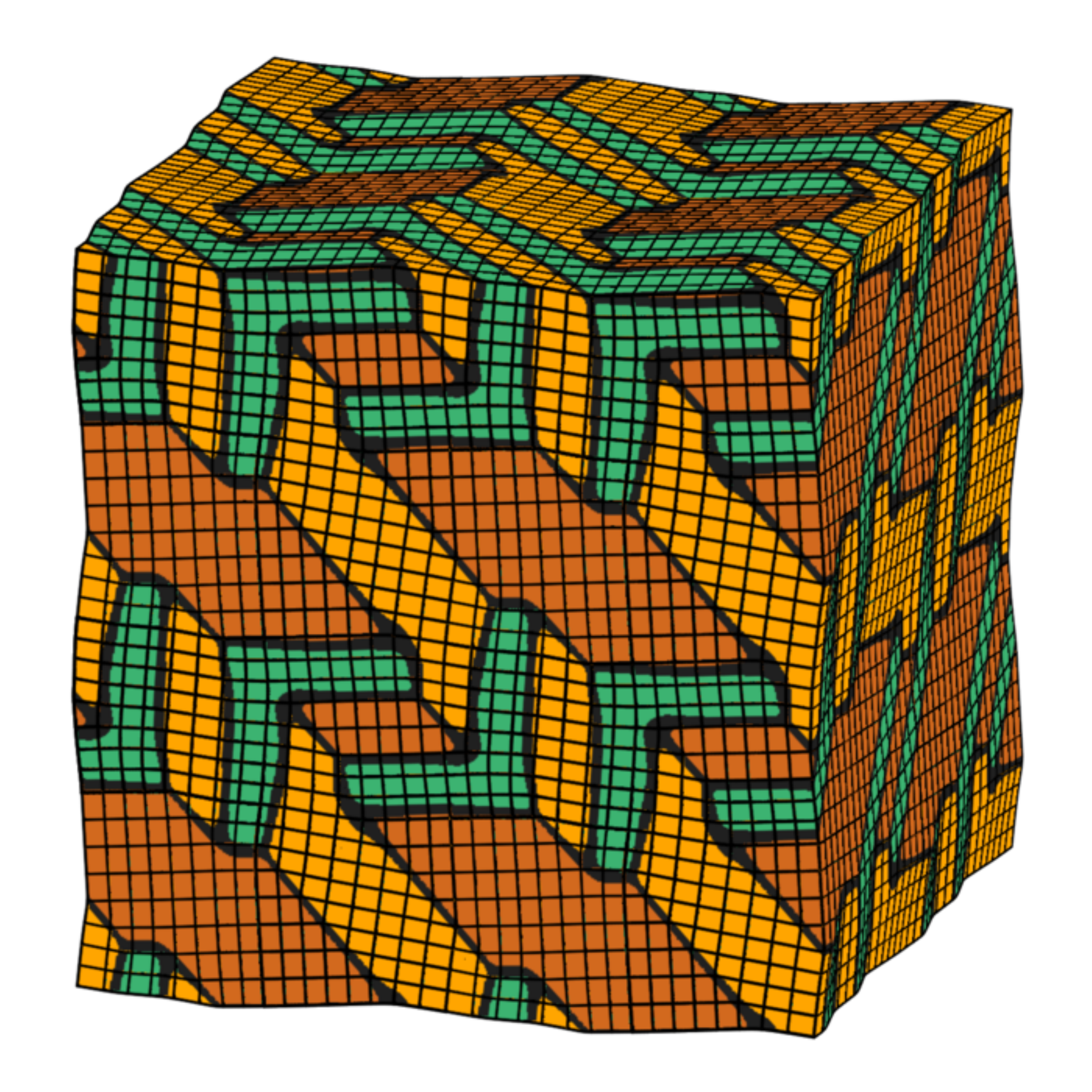}
        \caption{}
        \label{Fi:m19_222}
    \end{subfigure}
    ~
    \begin{subfigure}[b]{1.5in}
        \centering
        \includegraphics[scale=0.15]{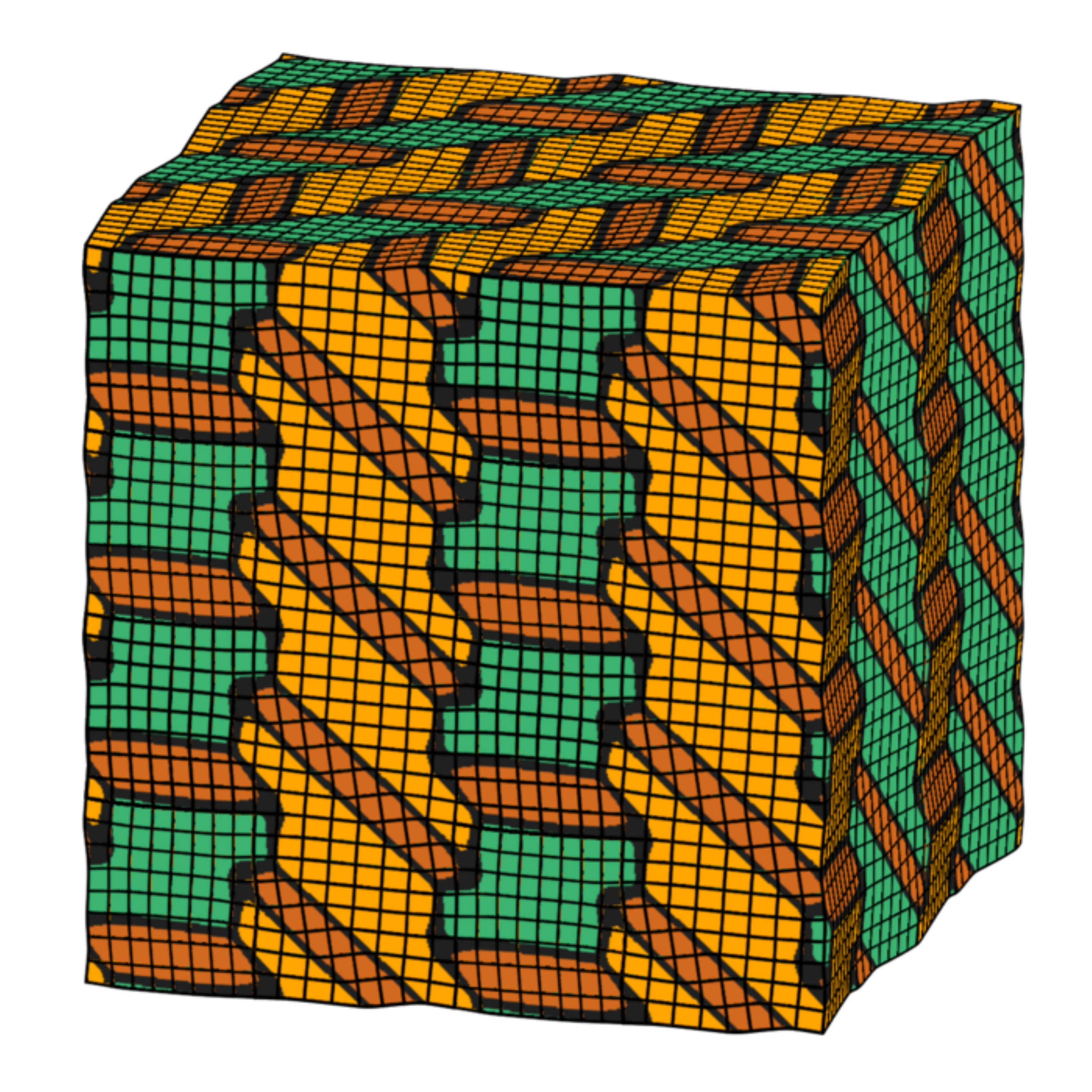}
        \caption{}
        \label{Fi:m28_222}
    \end{subfigure}
    ~
    \begin{subfigure}[b]{1.5in}
        \centering
        \includegraphics[scale=0.15]{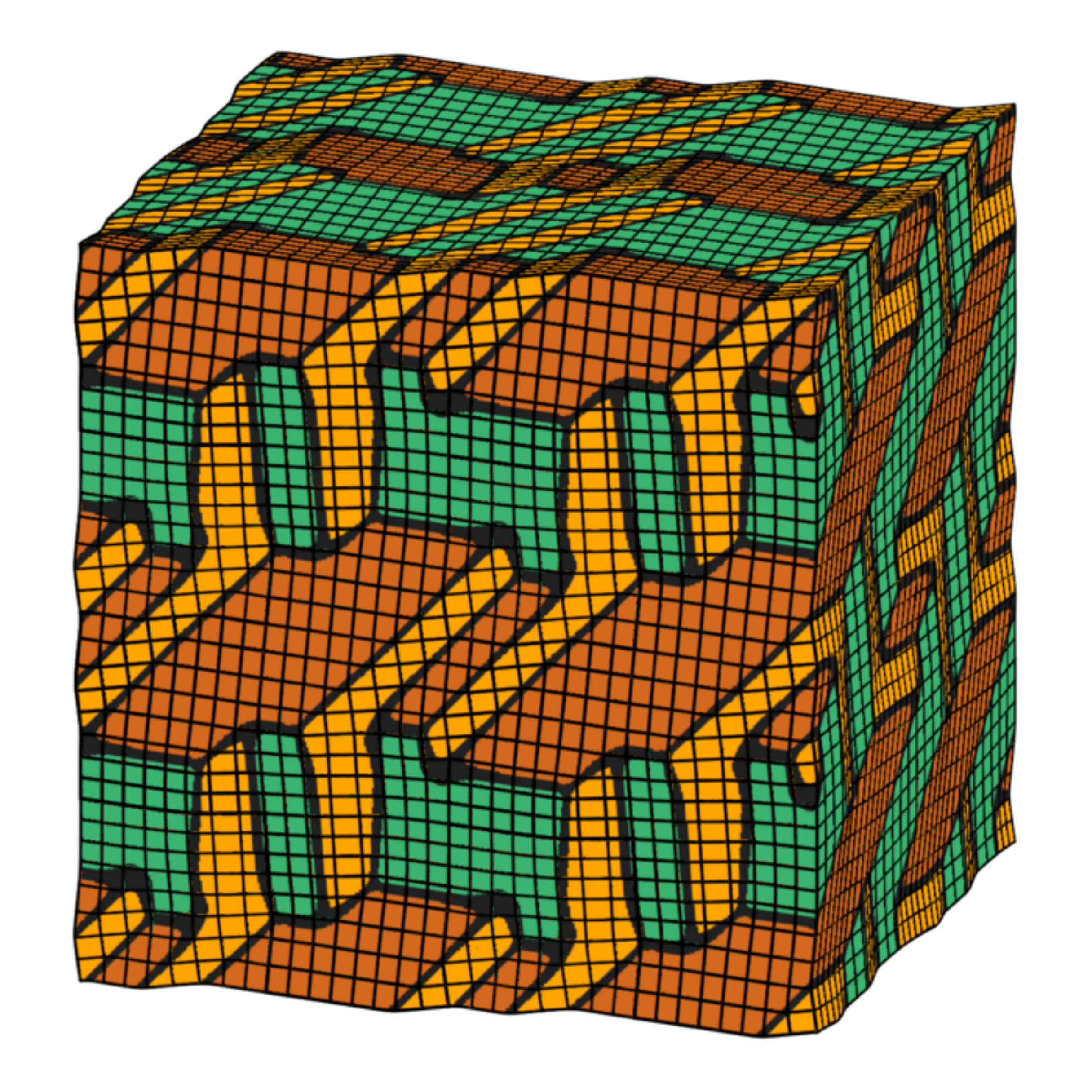}
        \caption{}
        \label{Fi:m20_222}
    \end{subfigure}
    
    \begin{subfigure}[b]{1.5in}
        \centering
        \includegraphics[scale=0.15]{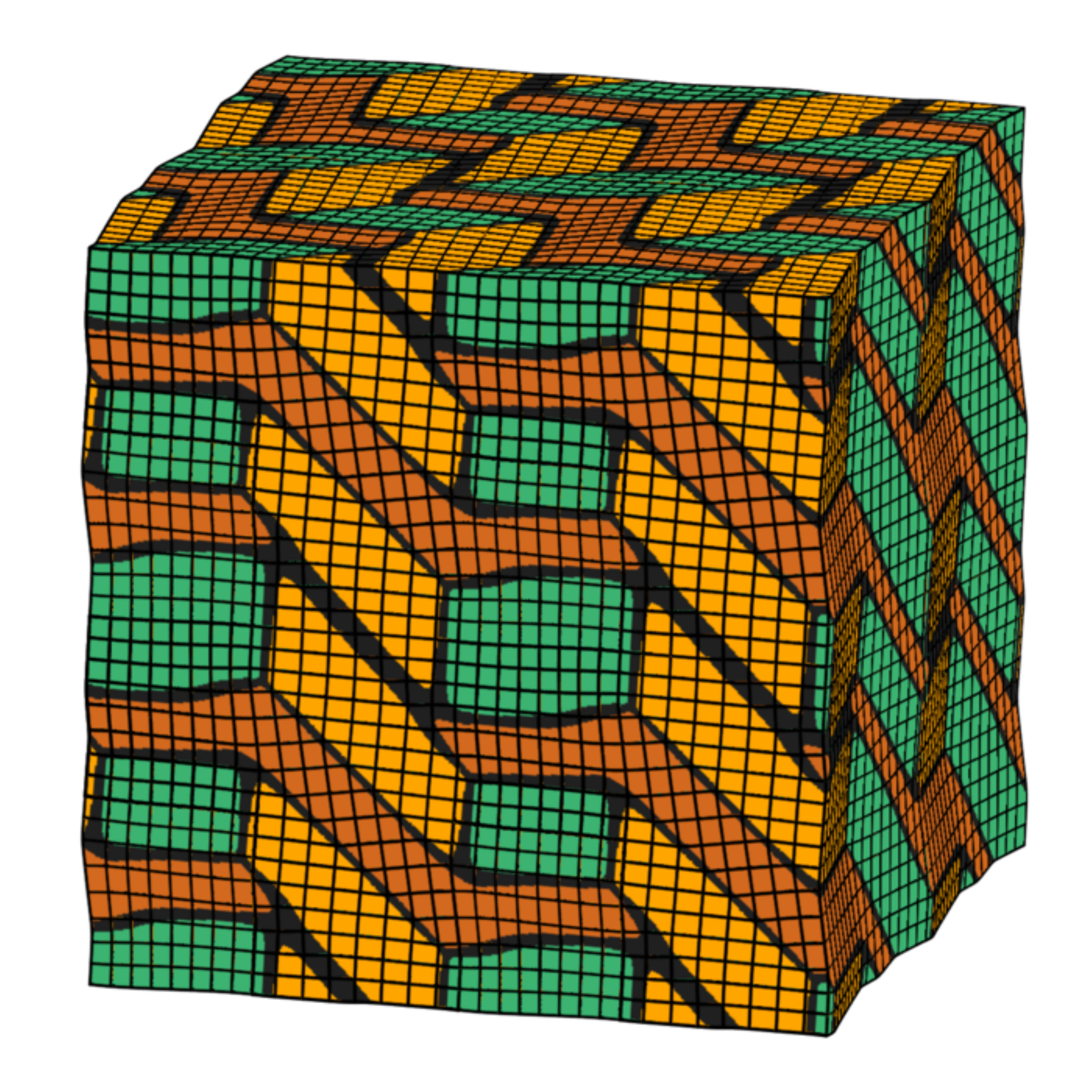}
        \caption{}
        \label{Fi:m42_222}
    \end{subfigure}
    ~
    \begin{subfigure}[b]{1.5in}
        \centering
        \includegraphics[scale=0.15]{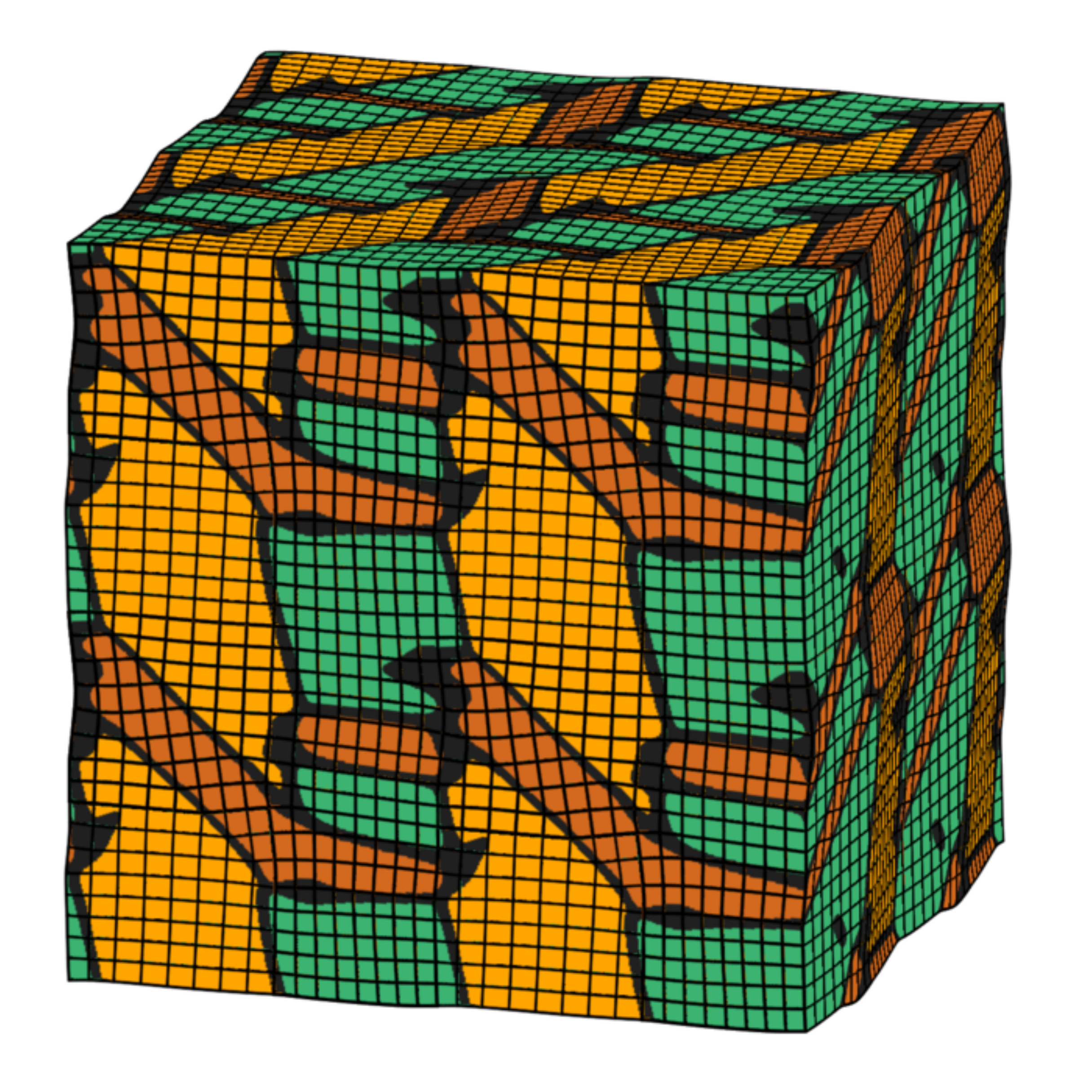}
        \caption{}
        \label{Fi:m34_222}
    \end{subfigure}
    ~
    \begin{subfigure}[b]{1.5in}
        \centering
        \includegraphics[scale=0.15]{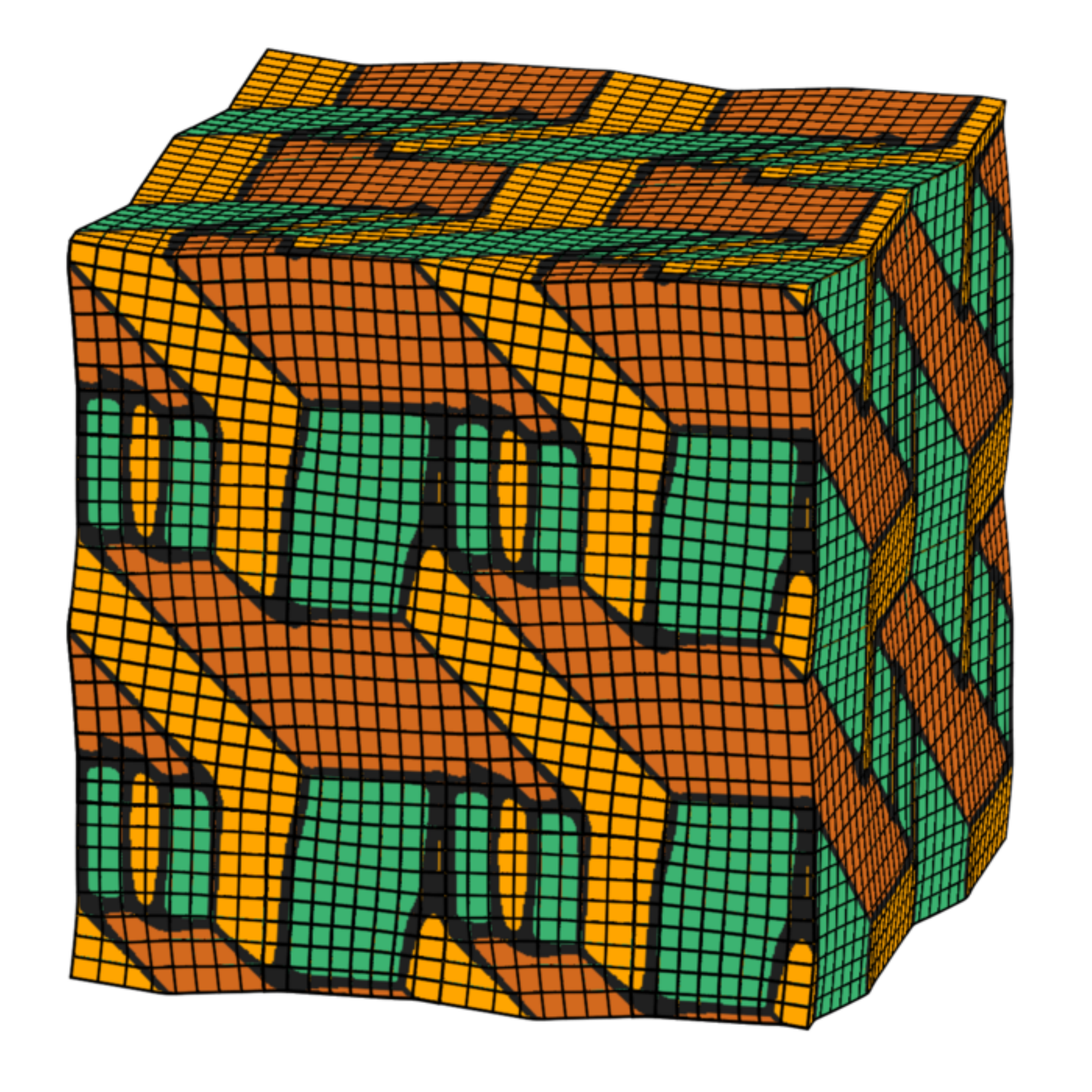}
        \caption{}
        \label{Fi:m8_222}
    \end{subfigure}
    ~
    \begin{subfigure}[b]{1.5in}
        \centering
        \includegraphics[scale=0.15]{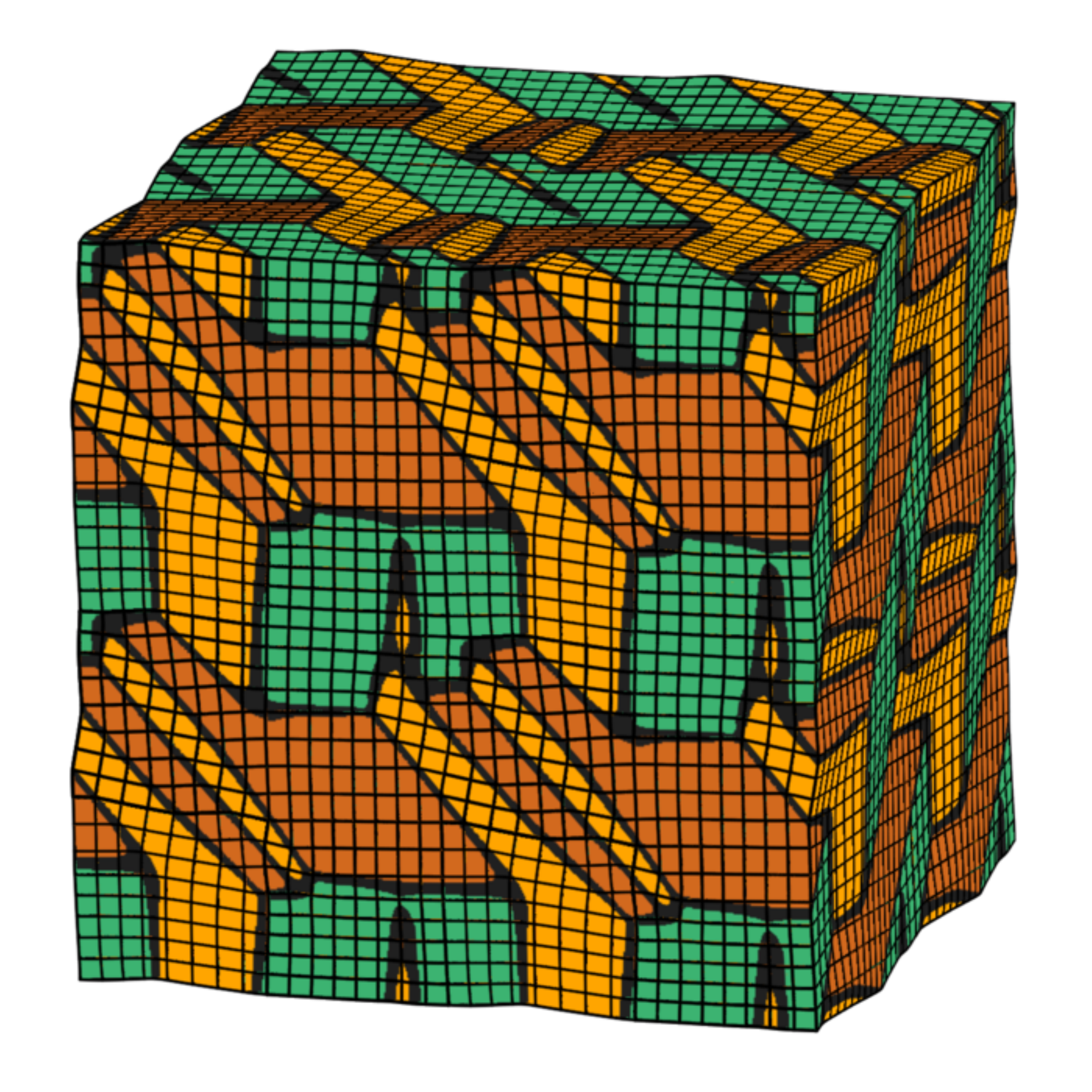}
        \caption{}
        \label{Fi:m27_222}
    \end{subfigure}
    \caption{A family of eight martensitic microstructures obtained from the gradient-coercified non-convex model of hyperelasticity.}
    \label{Fi:m_222}
\end{figure}

\begin{enumerate}
\item The data are obtained by DNS on a physical model that generates realistic martensitic microstructures by solving a high-dimensional problem of gradient elasticity at finite strains based on a non-convex hyperelastic free energy density function. In this regard, the homogenization we undertake is meaningful because the microstructures are obtained by solving the appropriate physical problem. We also note that much mathematical work has gone into analysis of such problems of martensitic microstructures. Our study adds a numerical perspective to this area.

\item From the standpoint of machine learning methods, we have confirmed that DNNs are able to represent the correct homogenized response that arises from complex microstructures.

\item Perhaps most interesting is that while being trained only on the scalar free energy densities, the DNNs deliver a high-fidelity representation that predicts the correct derivative fields, specifically the stress. Thus they recapitulate the hyperelastic response of these materials. The ability to recover derivative fields from machine learning models could prove to be of immense importance in every branch of physics. In particular it offers a path toward formal scale bridging where machine learning models could be trained on fine scale physics and correctly abstract their complexity to predict derivative fields at coarser scales, thus preserving the structure of the coarse-grained theories.

\item This investigation has restricted itself to a single martensitic microstructure: that shown in Figures \ref{Fi:m0_111} and \ref{Fi:m0_222}. A much more ambitious study awaits, where neural network representations are trained against a family of microstructures such as those appearing in Figure \ref{Fi:m_222}. This study will be the subject of a future communication.
\end{enumerate}

\section*{Acknowledgements}
\label{sec:acknowledgements}
We gratefully acknowledge the support of Toyota Research Institute, Award \#849910, ``Computational framework for data-driven, predictive, multi-scale and multi-physics modeling of battery materials". Simulations in this work were performed using the Extreme Science and Engineering Discovery Environment (XSEDE) Stampede2 at the Texas Advance Computing Center through allocations TG-MSS160003 and TG-DMR180072. XSEDE is supported by National Science Foundation grant number ACI-1548562.

\bibliographystyle{amsplain}
\bibliography{reference}

\end{document}